\documentclass[11pt,twocolumn]{aastex63}

\received{May 07, 2023}
\accepted{July 12, 2021}
\submitjournal{AJ}

\newcommand{\rh}{$r_{\rm h}$}

\begin{document}
\title{Binary Star Evolution in Different Environments: Filamentary, Fractal, Halo and Tidal-tail Clusters}

\shorttitle{Binary Evolution in different environments}

\shortauthors{Pang et al.}

\author[0000-0003-3389-2263]{Xiaoying Pang}
    \affiliation{Department of Physics, Xi'an Jiaotong-Liverpool University, 111 Ren’ai Road, Dushu Lake Science and Education Innovation District, Suzhou 215123, Jiangsu Province, P. R. China.}
    \email{Xiaoying.Pang@xjtlu.edu.cn}
    \affiliation{Shanghai Key Laboratory for Astrophysics, Shanghai Normal University, 
                100 Guilin Road, Shanghai 200234, P. R. China}
    
\author[0000-0002-7117-9533]{Yifan Wang}
    \affiliation{Department of Physics, Xi'an Jiaotong-Liverpool University, 111 Ren’ai Road, Dushu Lake Science and Education Innovation District, Suzhou 215123, Jiangsu Province, P. R. China.}
    
\author[0000-0003-4247-1401]{Shih-Yun Tang}
    \affiliation{Lowell Observatory, 1400 West Mars Hill Road, Flagstaff, AZ 86001, USA}
    \affiliation{Department of Physics and Astronomy, Rice University, 6100 Main Street, Houston, TX 77005, USA}

\author[0000-0001-5295-1682]{Yicheng Rui}
    \affiliation{Tsung-Dao Lee Institute, Shanghai Jiao Tong University, Shengrong Road 520, Shanghai, 201210, People's Republic Of China}
    \affiliation{School of Physics and Astronomy, Shanghai Jiao Tong University, 800 Dongchuan Road, Shanghai 200240, People's Republic of China}

\author[0000-0003-4247-1401]{Jing Bai}
    \affiliation{Department of Physics, Xi'an Jiaotong-Liverpool University, 111 Ren’ai Road, Dushu Lake Science and Education Innovation District, Suzhou 215123, Jiangsu Province, P. R. China.}

\author[0000-0003-4247-1401]{Chengyuan Li}
    \affiliation{School of Physics and Astronomy, Sun Yat-sen University, Daxue Road, Zhuhai, 519082, China}
    \affiliation{CSST Science Center for the Guangdong-Hong Kong-Macau Greater Bay Area, Zhuhai, 519082, China}

\author[0000-0001-6039-0555]{Fabo Feng}
    \affiliation{Tsung-Dao Lee Institute, Shanghai Jiao Tong University, Shengrong Road 520, Shanghai, 201210, People's Republic Of China}
    \affiliation{School of Physics and Astronomy, Shanghai Jiao Tong University, 800 Dongchuan Road, Shanghai 200240, People's Republic of China}

\author[0000-0002-1805-0570]{M.B.N. Kouwenhoven}
     \affiliation{Department of Physics, Xi'an Jiaotong-Liverpool University, 111 Ren’ai Road, Dushu Lake Science and Education Innovation District, Suzhou 215123, Jiangsu Province, P. R. China.}

\author[0000-0003-0262-272X]{Wen-Ping Chen}
\affiliation{Institute of Astronomy, National Central University, 
        300 Zhongda Road, Zhongli, Taoyuan 32001, Taiwan}

\author[0000-0001-6355-0673]{Rwei-ju Chuang}
\affiliation{Fu-Jen Catholic University, Institute of Applied Sciences, 510 Zhongzheng Rd., Xinzhuang Dist, New Taipei City 24205,Taiwan}


\begin{abstract} 

Using membership of 85 open clusters from previous studies \citep{pang2021a,pang2021b,pang2022a,li2021} based on Gaia DR\,3 data, we identify binary candidates in the color-magnitude diagram, for systems with mass ratio $q>0.4$.  
The binary fraction is corrected for incompleteness at different distances due to the Gaia angular resolution limit. We find a decreasing binary fraction with increasing cluster age, with substantial scatter. For clusters with a total mass $>$200 $M_\odot$, the binary fraction is independent of cluster mass.
The binary fraction depends strongly on stellar density. Among four types of cluster environments, the lowest-density filamentary and fractal stellar groups have the highest mean binary fraction: 23.6\% and 23.2\%, respectively.
The mean binary fraction in tidal-tail clusters is $20.8\%$, and is lowest in the densest halo-type clusters: 14.8\%. 
We find clear evidence of early disruptions of binary stars in the cluster sample.
The radial binary fraction depends strongly on the cluster-centric distance across all four types of environments, with the smallest binary fraction within the half-mass radius \rh, and increasing towards a few \rh. Only hints of mass segregation is found in the target clusters. The observed amount of mass segregation is not significant to generate a global effect inside the target clusters. We evaluate the bias of unresolved binary systems (assuming a primary mass of 1\,$M_\odot$) in 1D tangential velocity, which is 0.1--1$\,\rm km\,s^{-1}$. Further studies are required to characterize the internal star cluster kinematics using Gaia proper motions.

\end{abstract}


\section{Introduction}

Binary systems play a critical role in our understanding of the star formation process. The likelihood of a star being in a binary system is positively correlated with the mass of the star \citep{sana_binary_2012,raghavan_survey_2010,duchene_stellar_2013,offner_origin_2022}. Binary stars can originate from either fragmentation of protostellar cores, filaments, or accretion disks, or from dynamical interactions between stars \citep{moe_close_2019,moeckel2007,bate2012,offner_origin_2022}. The properties of binary stars exhibit significant variations depending on the specific conditions under which the star formation process occurs. To understand the conditions leading to different binary populations, it is therefore necessary to investigate the frequency of observed binaries across clustered environments of different density and morphology, ranging from dispersing filamentary stellar groups or associations, disrupted tidal tail clusters to dense spherical halo-type star clusters \citep[e.g.,][]{pang2022a}, to gain a better understanding of the star formation process.

Low-density stellar groups, such as associations, have a similar wide binary fraction as the field star population \citep[e.g.,][]{reipurth2007,kouwenhoven2005,kouwenhoven_primordial_2007,kraus_mapping_2011}. However, the binary fraction in dense open clusters, such as the Pleiades, the Praesepe cluster, and Alpha~Per, falls below that of the field \citep{deacon_wide_2020}. In such dense clustered environments, primordial wide binary systems are efficiently disrupted by stellar encounters, and are unable to survive until they become members of the field \citep[e.g.,][]{kroupa2001,portegies_zwart_star_2001,reipurth2007,parker2009fieldcluster}. The survival probability of a binary system depends on the binding energy of the system, the strength of the encounters, and the frequency of such encounters \citep{reipurth_multiplicity_2014}. In low-mass, loosely-bound clusters, a greater proportion of binary systems can survive the destruction process due to the lower encounter rate and the smaller mean kinetic energy of the colliding stars \citep{sollima_fraction_2010}.

According to the hierarchical formation scenario postulated by \citet{kruijssen2012}, regions of high stellar density are the birthplaces of dense, bound clusters. In contrast, loose filamentary groups of stars are frequently formed in environments with lower stellar densities \citep{pang2022a}. The observed binaries in the Galactic field result from a combination of relatively unprocessed binaries from loose stellar groups, and highly-processed binaries from dense clusters. Consequently, wide stellar binaries in the field are more likely to have originated from loose stellar groups, while close/intermediate binaries can originate from either source.

The density of the stellar environment plays a significant role in the evolution of binary systems. A correlation between the binary fraction and cluster density has been observed in globular clusters, with higher-density environments having lower binary fractions than their lower-density counterparts \citep{sollima_fraction_2007}. A similar correlation has been observed between the binary fraction and the cluster's integrated absolute magnitude, with brighter clusters having a smaller binary fraction. Brighter clusters have a higher number of stellar members and their stellar densities are generally higher. The rate at which close encounters occur is expected to be higher in massive clusters, which causes disruption of a large fraction of primordial binary systems \citep{milone2008}. However, this trend is not always evident in open clusters, particularly those younger than 100~Myr \citep{jaehnig_-sync_2017,kounkel_close_2019}. 

Besides stellar density, the age of a cluster has a significant impact on the spatial distribution of binary systems. When the cluster dynamically ages, the process of two-body relaxation leads to mass segregation, causing an increase in the binary fraction towards the center of globular clusters \citep{sollima_fraction_2007,milone_acs_2012}. This trend has also been found in $N$-body simulations \citep{portegies_zwart_star_2004,shu2021}. As a consequence of mass segregation, most massive binaries migrate to the cluster center. This process is accelerated when a young, substructured cluster experiences mergers \citep[e.g.,][]{allison2009}. Strong and frequent encounters in the center can cause binaries to harden, and can in some cases result in Roche-lobe overflow or coalescence, leading to the formation of blue stragglers, and cataclysmic variable stars \citep{hurley2001,pang2022b}. 
\cite{sollima_fraction_2007} find that in 13 low-density globular clusters, there is a clear negative correlation between binary fraction and age, suggesting that longer processing times increase the likelihood of binary destruction. However, when the sample is expanded to 59 clusters in \citet{milone_acs_2012}, there is no clear evidence to support such a correlation.

Open clusters are much younger than globular clusters and are therefore expected to have a higher binary fraction on average. However, no clear dependence of the binary fraction on cluster age was found in the five open clusters studied by \citet{sollima_fraction_2010}, which may be attributed to the limited sample size. Many studies have been conducted to examine the binary fraction within individual open clusters  \citep{cohen_bayesian_2019,jerabkova_when_2019,li_modeling_2020,jadhav_high_2021,malofeeva_unresolved_2022-1,malofeeva2023}. However, there are numerous examples in literature in which the measured binary fractions obtained for the same open cluster are different. These inconsistencies in the published properties of the binary population for the same cluster may be attributed to differences in cluster member identification and binary selection methods. To elucidate the relationship between binary fraction, age, and stellar density, a uniform analysis of a large sample of open clusters covering a wide range of ages and masses is required. Young open clusters are particularly useful for investigating the primordial binary content and the process of binary disruption in the early stages of cluster evolution. We will use a consistent reduction approach to identify binary content for 85 open clusters whose members have been determined via the same approach in \citet{pang2022a,pang2021a,pang2021b} and \citet{li2021}. By averaging the uniform binary data, we aim to obtain a comprehensive overview of binary evolution in different clustered environments.

This paper is organized as follows. In Section~\ref{sec:gaia}, the data and membership are introduced. In Section~\ref{sec:identify} we present the binary identification and the related uncertainties. We correct the binary fraction for incompleteness and derive the total binary fraction given different mass ratio profiles. We then discuss the dependence of binary fraction on cluster parameters in Section~\ref{sec:dependence}. In Section~\ref{sec:environ}, we investigate the evolution of the binary population in different clustered environments (i.e., different morphological types). The radial distributions of the binary systems in the representative clusters are presented in  Section~\ref{sec:morph_indi}. In Section~\ref{sec:mass-seg}, we present how mass segregation affects the representative clusters among the four morphological types, and we study its relation to binarity. In Section~\ref{sec:vel-disp}, the 1D velocity dispersion of clusters is  measured, with simulations to estimate the binaries' effect on the measured dispersion. We cross-match our results to previous studies in Section~\ref{sec:cross}. Finally, we provide a brief summary of our findings in Section~\ref{sec:summary}.

\section{Gaia data and cluster membership}\label{sec:gaia}

In this study, we utilize member stars from a sample of 85 open clusters as identified in previous works by \citet{pang2021a,pang2021b,pang2022a} and \citet{li2021}. Note that these four studies required member stars to have parallaxes and photometric measurements within 10\% uncertainties. When 
binary component separation close to the Gaia resolution limit, they will suffer from worse astrometric solutions than single stars. Hence these unresolved binary systems may be removed in afore-mentioned studies. 
The estimate of the unresolved binary systems in this work will be a lower limit due to this bias. However, it is a common disadvantage for the present-day investigations.
 The selection of members is performed with the aid of the machine learning algorithm \texttt{StarGo} \citep{yuan2018}, which makes use of the Gaia EDR\,3 data \citep{gaia2021}. The member stars are selected with a contamination rate of 5\% \citep[first developed in][]{pang2020}, resulting in a corresponding membership probability of 95\%.

All stars present in Gaia EDR\,3, with solutions, should be treated as single stars, with no other sources within the minimum angular resolution. The photometry used in this study is based on two-parameter solutions. For neighboring sources that are distinctly separate and have an angular separation of 0.18--0.6~arcsec, only two-parameter solutions were available  \citep{lindegren_gaia_2021}. In Figure~6 of \citet{lindegren_gaia_2021}, the angular separation for sources with a $G$ magnitude of 15 mag is about 0.6~arcsec. In this study, we treat the 0.6~arcsec as the minimum angular resolution for Gaia data, beyond which a single star can be distinguished. Any stars with neighbors closer than 0.6~arcsec are deemed unresolved binaries.

Following the correction of G-band photometry in Gaia DR\,3 by \citet{riello2021}, the uncertainty in the photometry of sources with magnitudes fainter than $G=13$\,mag has been improved. In order to obtain more accurate photometric and kinematic data for each star, we cross-match the members of these 85 clusters with Gaia DR\,3 \citep{gaia_collaboration_gaia_2022}. Our analysis subsequently relies on the data from Gaia DR\,3.


\section{Binary identification}\label{sec:identify}

\subsection{Binary Selection Through the Color-Magnitude Diagram}\label{sec:select}

Unresolved binary systems with different mass ratios exhibit distinct features in the color-magnitude diagram (CMD). Through this study, we define the mass ratio as $q=M_2/M_1$, where $M_1$ is the mass of the primary star and $M_2$ is the mass of the secondary star ($M_2\leq M_1$). Unresolved binary systems appear redder and brighter than the main sequence (MS) in the CMD. When the binary components are of equal mass $q=1$, they form a distinctive ``binary sequence'' that is parallel to the MS, and is 0.75 magnitudes brighter. Unresolved binaries with high $q$ (i.e., $q>0.7$) are located at a significant distance from the MS ridge-line and are therefore relatively easy to detect. On the other hand, binaries with small $q$ are located near the MS ridge-line, making them difficult to distinguish from single MS stars.

\begin{figure*}[tb!]
\centering
\includegraphics[angle=0, width=1.\textwidth]{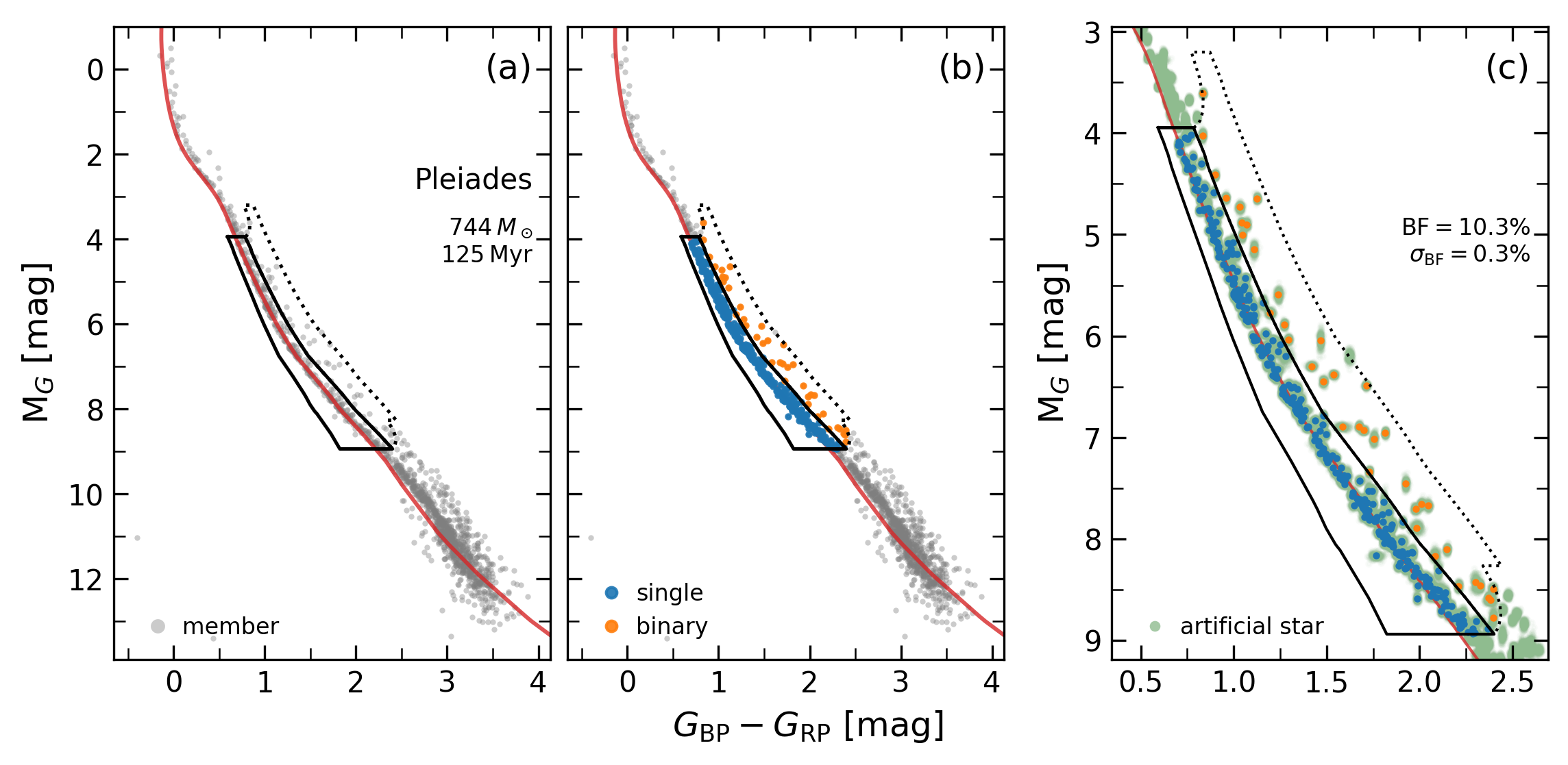}
    \caption{Binary identification via 
    the color-magnitude diagram (CMD). The Pleiades cluster is shown as an example. (a): best-fitted isochrone for the Pleiades (red curve). The region enclosed by the solid curve demarcates the single-star boundary defined in this work, whereas the dotted curve identifies the binary region. The right boundary of solid polygon is the binary isochrone of $q=0.4$, which is the boundary that we use to separate single and binary stars. The upper limit of the solid boundary is $M_G=4$\,mag, and the lower limit is 0.5$M_\odot$ ($M_G=9-10$\,mag). Members of the Pleiades are over-plotted (grey circles). (b): Stars within the solid-curve boundary are selected as single stars, and are highlighted as blue dots. Binary stars (orange dots) are stars located within the binary boundary (dashed curves). (c): A zoom-in for the binary selected region. The green dots in the background are artificial stars produced via Monte Carlo simulations based on the photometric uncertainty (see error estimation in Section~\ref{sec:err} for details). 
    }
\label{fig:CMD}
\end{figure*}

To select binary candidates, we identify stars located on the red side of the MS ridge-line. In this study, we adopt the best-fitted PARSEC isochrones \citep{{bressan2012,chen2015}} from \citet{pang2021a,pang2021b,pang2022a} and \citet{li2021} to represent the MS ridge-line, since it also provides information about stellar masses and cluster age. The stellar mass will be used to determine the binary-single boundary, and the cluster age is a key parameter in the binary correlation investigation (see Section~\ref{sec:bf_age}). Moreover, the deviation of the best-fit isochrone from the cluster ridge line occurs mainly for the faint stars (e.g., M dwarfs), which are below our binary selection region (solid and dotted polygon in Figure~\ref{fig:CMD}) and therefore will not affect the binary identification in this work. The systematic error for all clusters from isochrone fitting is identical, so that the investigated relations of all clusters, e.g., binary fraction and other variables, will not be influenced by the choice of the isochrone.  Considering the negligible shifts of unresolved binaries with small $q$ from the MS ridge-line, we only study binary systems with $q>q_0$, where $q_0$ is a certain minimum, so we can identify binary cluster members with high confidence. To accomplish this, we adopt the approach from \citet{milone_acs_2012}, and identify unresolved binary stars with higher mass ratios $q>0.4$ (see Figure~\ref{fig:CMD}).

 We set the binary isochrone of $q_0=0.4$ as the boundary between single and binary stars (right boundary of the solid polygon in Figure~1). We compute the binary isochrone by treating the two components in the binary system as unresolved. Any stars at locations that are both brighter and redder than this  binary isochrone ($q_0=0.4$) are considered as unresolved binary stars or multiple systems. Throughout this study, we do not consider the possibility of triple systems or higher-order multiples. The difference between the best fitted single isochrone and the binary isochrone of $q_0=0.4$ is considered as the typical width of the single region. 
On the left side of the best-fitting isochrone, We establish the left boundary of the single star region as the 3-5 times the typical width, depending on the photometric uncertainty. To establish the upper limit of the binary region, we adopt the fiducial binary isochrone for $q=1$, which is 0.75\,mag brighter than the single-star isochrone. Additionally, we add a photometric uncertainty of 0.1\,mag to the $q=1$ binary isochrone to be the right boundary of the binary region. 

When a pre-main-sequence (PMS) star rotates fast  \citep{li_discovery_2017,cordoni2018,li_extended_2019}, it will appear redder than the MS stars, and resembles an unresolved binary system. \citet{cordoni2018} have shown that nearly all star clusters exhibit a substantial amount of MS broadening due to rotating MS stars with an absolute G magnitude of less than 4 mag. The latest study by \citet{he2022} suggests that binary systems of mass ratio less than 0.6 are indistinguishable from rapidly rotating stars based on their colors. 
On the other hand, circumstellar disks around  PMS stars, such as young class~II objects, can make them appear redder compared to the disk-less class~III objects. These young stellar objects have a typical spectral type of M1 \citep[e.g.,][]{lopez-valdivia_igrins_2021}, where the effective temperature is estimated to range from roughly 3600\,K \citep{pecaut_intrinsic_2013} to 3700\,K \citep{herczeg_optical_2014}. According to the PMS evolutionary track by \citet{baraffe2015}, at the age of 3~Myr (our youngest cluster), the effective temperature 3600-3700\,K correspond to a stellar mass of 0.5--0.7 solar masses. Therefore, to avoid contamination due to accretion disks and fast rotation, we limit the selection of single and binary stars to those with a stellar mass greater than $\rm 0.5\,M_\odot$ (corresponding to $M_G=9-10$\,mag in clusters of different ages) and an absolute $G$ magnitude fainter than $M_{G}=4$\,mag (the upper and lower boundaries of both the single and binary boundary, i.e., the regions outlined by the solid and dotted curves in Figure~\ref{fig:CMD} (a), (b), and (c)). 

The binary candidates are therefore mainly FGK stars. Any observed members outside the two regions described above are considered outliers and are excluded when calculating the binary fraction. Figure~\ref{fig:CMD} illustrates an example of the single and binary star regions we defined for the Pleiades cluster.

In this study, we quantify the binary fraction of systems with a mass ratio $q>q_0=0.4$. This is achieved by computing the ratio between the total number of binary systems and the overall number of stars within the selected regions: 
\begin{equation}\label{eq:binary_frac}
    f_{bin}^{q>0.4}=\frac{N_{bin}}{(N_{sin}+N_{bin})}
\end{equation}
where $N_{bin}$ and $N_{sin}$ are the number of binary systems and single stars, respectively. The fraction of binaries with $q>0.4$, $f^{q>0.4}_{bin}$, for all 85 open clusters is presented in the column~5 of Table~\ref{tab:bin_frac}. In total, 1811 binary candidates are identified among target clusters. Only two of them are cross-matched with the non-single star catalog published by Gaia DR\,3  \citep{gaia_collaboration_gaia_2022} that was determined via spectra.

The accuracy of our binary content analysis can be affected by contamination from field stars, particularly for faint stars with masses below 0.5\,$M_\odot$. The global contamination rate of member stars is limited to 5\% \citep{pang2022a}, which mainly affects stars fainter than the binary selection regions. The field star contamination in the binary selection region (as shown in Figure~\ref{fig:CMD}) is even lower then 5\%. Therefore, our binary candidates are statistically robust. The selection of binaries in the CMD is advantageous in efficiency and independent of influence by the orbital period or inclination angles of the binary systems \citep{milone_acs_2012}. Most of our target clusters are consistent with a uniform extinction; we adopt the best-fitting extinction value from isochrone fitting in \citet{pang2021a,pang2021b,pang2022a} and \citet{li2021} for members of each cluster. 


\subsection{Error Estimation}\label{sec:err}

The mean photometric uncertainty of Gaia DR\,3 is about 1\,mmag \citep{gaia_collaboration_gaia_2022}, which may result in uncertain placement of cluster members in the CMD and may lead to variations in the number of measured single or binary stars (as shown by the grey dots in Figure~\ref{fig:CMD} (c)). To account for this effect, we simulate the impact of observational photometric uncertainty by assigning each member a new magnitude drawn from a Gaussian distribution centered at the observed value and with a dispersion of the photometric uncertainty of the individual star. With the updated locations of each member in the CMD, we then measure the number of stars within the single and binary regions again and recalculate the binary fraction using Equation~\ref{eq:binary_frac}. This procedure is repeated 2000 times, and the average difference between the newly obtained binary fraction and the original one from Section~\ref{sec:select} is considered as the associated error of the binary fraction $f_{bin}^{q>0.4}$.

\startlongtable
\begin{deluxetable*}{L RRR R RRR R}
	\tablecaption{Binary fraction of 85 open clusters.
	\label{tab:bin_frac}}
	\tabletypesize{\scriptsize}
	\tablehead{
	    \colhead{Cluster}           & 
		\colhead{$M_{\rm cl}$}      & 
        \colhead{Age}               & 
        \colhead{Dist.}             &
		\colhead{$f^{q>0.4}_{bin}$}   &
        \colhead{$f^{q>0.4}_{binc}$}  &
        \colhead{$f^{tot}_{k07}$}    &
		\colhead{$f^{tot}_{f05}$}     &
		\colhead{$f^{tot}_{uni}$}  
		\\
		\colhead{}              & 
		\colhead{($M_\odot$)}   & 
		\colhead{(Myr)}         &
		\colhead{(pc)}          &
		\colhead{(\%)}          &
  	    \colhead{(\%)}          &
		\colhead{(\%)}          &
		\colhead{(\%)}          &
		\colhead{(\%)}     
		\\
	    \cline{1-9}
	    \colhead{(1)}   & 
	    \colhead{(2)}   & \colhead{(3)} & 
	    \colhead{(4)}   & \colhead{(5)} &
	    \colhead{(6)}   & \colhead{(7)} & 
	    \colhead{(8)}   & \colhead{(9)}
		}
	\startdata
\mathrm{ASCC\ 16} & 241 & 10 & 346.4 & 11.8\pm1.0 & 17.6\pm1.4 & 41.9\pm3.3 & 26.3\pm2.1 & 29.3\pm2.3\\
\mathrm{ASCC\ 19} & 197 & 8 & 354.7 & 13.3\pm1.2 & 19.0\pm1.8 & 45.2\pm4.3 & 28.4\pm2.7 & 31.7\pm3.0\\
\mathrm{ASCC\ 58} & 254 & 52 & 477.1 & 14.0\pm1.2 & 19.4\pm1.9 & 46.2\pm4.5 & 29.0\pm2.8 & 32.3\pm3.2\\
\mathrm{ASCC\ 105} & 67 & 73 & 557.8 & 10.0\pm1.9 & 13.5\pm2.9 & 32.1\pm6.9 & 20.1\pm4.3 & 22.5\pm4.8\\
\mathrm{ASCC\ 127} & 182 & 14 & 374.4 & 14.5\pm1.6 & 20.8\pm2.1 & 49.5\pm5.0 & 31.0\pm3.1 & 34.7\pm3.5\\
\mathrm{Alessi\ 3} & 123 & 631 & 277.3 & 20.6\pm0.6 & 30.8\pm0.8 & 73.3\pm1.9 & 46.0\pm1.2 & 51.3\pm1.3\\
\mathrm{Alessi\ 5} & 242 & 52 & 397.7 & 11.7\pm0.8 & 16.7\pm1.1 & 39.8\pm2.6 & 24.9\pm1.6 & 27.8\pm1.8\\
\mathrm{Alessi\ 9} & 61 & 265 & 207.6 & 23.8\pm0.6 & 38.1\pm0.8 & 90.7\pm1.9 & 56.9\pm1.2 & 63.5\pm1.3\\
\mathrm{Alessi\ 20} & 241 & 8 & 423.4 & 20.4\pm2.1 & 29.2\pm3.0 & 69.5\pm7.1 & 43.6\pm4.5 & 48.7\pm5.0\\
\mathrm{Alessi\ 20\ gp1} & 172 & 11 & 411.9 & 21.9\pm1.7 & 31.4\pm2.4 & 74.8\pm5.7 & 46.9\pm3.6 & 52.3\pm4.0\\
\mathrm{Alessi\ 20\ isl1} & 97 & 100 & 458.9 & 10.9\pm1.4 & 15.1\pm2.0 & 36.0\pm4.8 & 22.5\pm3.0 & 25.2\pm3.3\\
\mathrm{Alessi\ 24} & 119 & 87 & 484.4 & 14.5\pm6.2 & 20.1\pm8.3 & 47.9\pm19.8 & 30.0\pm12.4 & 33.5\pm13.8\\
\mathrm{Alessi\ 62} & 143 & 692 & 618.9 & 21.3\pm3.7 & 28.9\pm4.8 & 68.8\pm11.4 & 43.1\pm7.2 & 48.2\pm8.0\\
\mathrm{BH\ 99} & 565 & 81 & 446.9 & 11.8\pm1.9 & 16.9\pm2.7 & 40.2\pm6.4 & 25.2\pm4.0 & 28.2\pm4.5\\
\mathrm{BH\ 164} & 196 & 64 & 420.2 & 12.9\pm4.1 & 18.5\pm5.6 & 44.0\pm13.3 & 27.6\pm8.4 & 30.8\pm9.3\\
\mathrm{Blanco\ 1} & 338 & 100 & 236.8 & 11.0\pm0.6 & 17.6\pm0.9 & 41.9\pm2.1 & 26.3\pm1.3 & 29.3\pm1.5\\
\mathrm{Collinder\ 69} & 401 & 13 & 398.9 & 33.9\pm1.4 & 48.5\pm2.1 & 115.5\pm5.0 & 72.4\pm3.1 & 80.8\pm3.5\\
\mathrm{Collinder\ 135} & 251 & 40 & 303.2 & 18.3\pm1.0 & 27.4\pm1.5 & 65.2\pm3.6 & 40.9\pm2.2 & 45.7\pm2.5\\
\mathrm{Collinder\ 140} & 179 & 49 & 384.5 & 13.2\pm1.3 & 18.9\pm1.8 & 45.0\pm4.3 & 28.2\pm2.7 & 31.5\pm3.0\\
\mathrm{Collinder\ 350} & 149 & 589 & 367.9 & 28.0\pm2.1 & 40.1\pm2.8 & 95.5\pm6.7 & 59.9\pm4.2 & 66.8\pm4.7\\
\mathrm{Coma\ Berenices} & 100 & 700 & 86.0 & 4.2\pm0.2 & 7.7\pm0.5 & 18.3\pm1.2 & 11.5\pm0.7 & 12.8\pm0.8\\
\mathrm{Group\ X} & 98 & 400 & 99.6 & 8.9\pm0.4 & 16.3\pm0.8 & 38.8\pm1.9 & 24.3\pm1.2 & 27.2\pm1.3\\
\mathrm{Gulliver\ 6} & 169 & 7 & 413.3 & 16.9\pm1.5 & 24.2\pm2.2 & 57.6\pm5.2 & 36.1\pm3.3 & 40.3\pm3.7\\
\mathrm{Gulliver\ 21} & 83 & 274 & 652.5 & 15.7\pm2.6 & 21.3\pm3.7 & 50.7\pm8.8 & 31.8\pm5.5 & 35.5\pm6.2\\
\mathrm{Huluwa\ 1} & 724 & 12 & 355.1 & 16.3\pm0.5 & 23.3\pm0.8 & 55.5\pm1.9 & 34.8\pm1.2 & 38.8\pm1.3\\
\mathrm{Huluwa\ 2} & 469 & 11 & 399.0 & 17.3\pm0.9 & 24.8\pm1.2 & 59.0\pm2.9 & 37.0\pm1.8 & 41.3\pm2.0\\
\mathrm{Huluwa\ 3} & 371 & 10 & 394.6 & 15.4\pm0.9 & 22.0\pm1.2 & 52.4\pm2.9 & 32.8\pm1.8 & 36.7\pm2.0\\
\mathrm{Huluwa\ 4} & 181 & 10 & 342.1 & 19.4\pm0.8 & 29.0\pm1.2 & 69.0\pm2.9 & 43.3\pm1.8 & 48.3\pm2.0\\
\mathrm{Huluwa\ 5} & 60 & 7 & 354.6 & 4.0\pm1.8 & 5.7\pm2.4 & 13.6\pm5.7 & 8.5\pm3.6 & 9.5\pm4.0\\
\mathrm{IC\ 348} & 142 & 4 & 316.5 & 30.3\pm1.9 & 45.3\pm2.8 & 107.9\pm6.7 & 67.6\pm4.2 & 75.5\pm4.7\\
\mathrm{IC\ 2391} & 138 & 49 & 151.3 & 9.7\pm0.7 & 15.5\pm1.0 & 36.9\pm2.4 & 23.1\pm1.5 & 25.8\pm1.7\\
\mathrm{IC\ 2602} & 187 & 44 & 151.4 & 8.5\pm0.6 & 13.6\pm1.0 & 32.4\pm2.4 & 20.3\pm1.5 & 22.7\pm1.7\\
\mathrm{IC\ 4665} & 159 & 35 & 347.2 & 13.0\pm1.2 & 19.4\pm1.9 & 46.2\pm4.5 & 29.0\pm2.8 & 32.3\pm3.2\\
\mathrm{IC\ 4756} & 509 & 954 & 473.7 & 16.6\pm1.0 & 23.0\pm1.3 & 54.8\pm3.1 & 34.3\pm1.9 & 38.3\pm2.2\\
\mathrm{LP\ 2371} & 81 & 19 & 366.9 & 12.5\pm2.3 & 17.9\pm3.7 & 42.6\pm8.8 & 26.7\pm5.5 & 29.8\pm6.2\\
\mathrm{LP\ 2373} & 98 & 4 & 386.9 & 26.3\pm3.2 & 37.7\pm4.3 & 89.8\pm10.2 & 56.3\pm6.4 & 62.8\pm7.2\\
\mathrm{LP\ 2373\ gp1} & 186 & 10 & 335.6 & 18.2\pm1.2 & 27.2\pm1.8 & 64.8\pm4.3 & 40.6\pm2.7 & 45.3\pm3.0\\
\mathrm{LP\ 2373\ gp2} & 544 & 8 & 349.3 & 15.4\pm0.9 & 23.0\pm1.4 & 54.8\pm3.3 & 34.3\pm2.1 & 38.3\pm2.3\\
\mathrm{LP\ 2373\ gp3} & 111 & 6 & 349.0 & 11.9\pm2.5 & 17.8\pm3.9 & 42.4\pm9.3 & 26.6\pm5.8 & 29.7\pm6.5\\
\mathrm{LP\ 2373\ gp4} & 295 & 6 & 363.0 & 17.6\pm0.9 & 25.2\pm1.4 & 60.0\pm3.3 & 37.6\pm2.1 & 42.0\pm2.3\\
\mathrm{LP\ 2383} & 277 & 49 & 363.5 & 12.1\pm1.1 & 17.3\pm1.6 & 41.2\pm3.8 & 25.8\pm2.4 & 28.8\pm2.7\\
\mathrm{LP\ 2388} & 149 & 21 & 497.3 & 34.5\pm2.6 & 47.9\pm3.6 & 114.0\pm8.6 & 71.5\pm5.4 & 79.8\pm6.0\\
\mathrm{LP\ 2428} & 112 & 200 & 436.0 & 37.4\pm1.8 & 53.5\pm2.6 & 127.4\pm6.2 & 79.9\pm3.9 & 89.2\pm4.3\\
\mathrm{LP\ 2429} & 148 & 1150 & 479.9 & 18.5\pm1.0 & 25.7\pm1.3 & 61.2\pm3.1 & 38.4\pm1.9 & 42.8\pm2.2\\
\mathrm{LP\ 2439} & 141 & 24 & 283.7 & 11.4\pm0.8 & 17.0\pm1.3 & 40.5\pm3.1 & 25.4\pm1.9 & 28.3\pm2.2\\
\mathrm{LP\ 2441} & 187 & 74 & 280.6 & 18.5\pm1.0 & 27.7\pm1.6 & 66.0\pm3.8 & 41.3\pm2.4 & 46.2\pm2.7\\
\mathrm{LP\ 2442} & 318 & 14 & 176.2 & 10.4\pm0.5 & 16.6\pm1.0 & 39.5\pm2.4 & 24.8\pm1.5 & 27.7\pm1.7\\
\mathrm{LP\ 2442\ gp1} & 111 & 8 & 139.2 & 14.8\pm1.9 & 27.0\pm3.6 & 64.3\pm8.6 & 40.3\pm5.4 & 45.0\pm6.0\\
\mathrm{LP\ 2442\ gp2} & 151 & 8 & 140.5 & 24.1\pm2.0 & 44.0\pm4.0 & 104.8\pm9.5 & 65.7\pm6.0 & 73.3\pm6.7\\
\mathrm{LP\ 2442\ gp3} & 64 & 8 & 142.4 & 13.8\pm2.4 & 25.2\pm4.5 & 60.0\pm10.7 & 37.6\pm6.7 & 42.0\pm7.5\\
\mathrm{LP\ 2442\ gp4} & 113 & 8 & 153.7 & 22.2\pm1.8 & 35.5\pm2.9 & 84.5\pm6.9 & 53.0\pm4.3 & 59.2\pm4.8\\
\mathrm{LP\ 2442\ gp5} & 76 & 8 & 154.1 & 12.5\pm0.1 & 20.0\pm0.2 & 47.6\pm0.5 & 29.9\pm0.3 & 33.3\pm0.3\\
\mathrm{Mamajek\ 4} & 282 & 371 & 449.9 & 9.9\pm1.6 & 14.2\pm2.3 & 33.8\pm5.5 & 21.2\pm3.4 & 23.7\pm3.8\\
\mathrm{NGC\ 1901} & 124 & 850 & 418.3 & 32.9\pm1.4 & 47.1\pm2.0 & 112.1\pm4.8 & 70.3\pm3.0 & 78.5\pm3.3\\
\mathrm{NGC\ 1977} & 108 & 3 & 392.3 & 16.1\pm1.7 & 23.0\pm2.5 & 54.8\pm6.0 & 34.3\pm3.7 & 38.3\pm4.2\\
\mathrm{NGC\ 1980} & 757 & 5 & 383.6 & 22.0\pm0.8 & 31.5\pm1.2 & 75.0\pm2.9 & 47.0\pm1.8 & 52.5\pm2.0\\
\mathrm{NGC\ 2232} & 205 & 24 & 319.6 & 13.1\pm1.0 & 19.6\pm1.5 & 46.7\pm3.6 & 29.3\pm2.2 & 32.7\pm2.5\\
\mathrm{NGC\ 2422} & 477 & 72 & 476.4 & 4.2\pm0.5 & 5.8\pm0.7 & 13.8\pm1.7 & 8.7\pm1.0 & 9.7\pm1.2\\
\mathrm{NGC\ 2451A} & 178 & 57 & 192.5 & 16.2\pm0.9 & 25.9\pm1.5 & 61.7\pm3.6 & 38.7\pm2.2 & 43.2\pm2.5\\
\mathrm{NGC\ 2451B} & 276 & 49 & 363.3 & 15.9\pm1.1 & 22.8\pm1.6 & 54.3\pm3.8 & 34.0\pm2.4 & 38.0\pm2.7\\
\mathrm{NGC\ 2516} & 1984 & 123 & 410.6 & 12.4\pm0.3 & 17.8\pm0.5 & 42.4\pm1.2 & 26.6\pm0.7 & 29.7\pm0.8\\
\mathrm{NGC\ 2547} & 302 & 39 & 387.1 & 14.0\pm1.3 & 20.0\pm1.7 & 47.6\pm4.0 & 29.9\pm2.5 & 33.3\pm2.8\\
\mathrm{NGC\ 3228} & 84 & 62 & 482.2 & 34.1\pm4.2 & 47.3\pm6.2 & 112.6\pm14.8 & 70.6\pm9.3 & 78.8\pm10.3\\
\mathrm{NGC\ 3532} & 2228 & 397 & 478.3 & 11.4\pm1.2 & 15.8\pm1.8 & 37.6\pm4.3 & 23.6\pm2.7 & 26.3\pm3.0\\
\mathrm{NGC\ 6405} & 598 & 79 & 457.6 & 15.8\pm1.8 & 21.9\pm2.6 & 52.1\pm6.2 & 32.7\pm3.9 & 36.5\pm4.3\\
\mathrm{NGC\ 6475} & 1023 & 185 & 279.5 & 9.3\pm0.4 & 13.9\pm0.5 & 33.1\pm1.2 & 20.7\pm0.7 & 23.2\pm0.8\\
\mathrm{NGC\ 6633} & 338 & 426 & 393.8 & 12.6\pm1.0 & 18.0\pm1.3 & 42.9\pm3.1 & 26.9\pm1.9 & 30.0\pm2.2\\
\mathrm{NGC\ 6774} & 152 & 2650 & 306.1 & 15.9\pm1.3 & 23.8\pm1.8 & 56.7\pm4.3 & 35.5\pm2.7 & 39.7\pm3.0\\
\mathrm{NGC\ 7058} & 126 & 80 & 365.2 & 10.4\pm1.5 & 14.9\pm2.1 & 35.5\pm5.0 & 22.2\pm3.1 & 24.8\pm3.5\\
\mathrm{NGC\ 7092} & 191 & 350 & 297.1 & 8.5\pm1.0 & 12.7\pm1.4 & 30.2\pm3.3 & 19.0\pm2.1 & 21.2\pm2.3\\
\mathrm{Pleiades} & 740 & 124 & 135.9 & 10.3\pm0.3 & 18.8\pm0.5 & 44.8\pm1.2 & 28.1\pm0.7 & 31.3\pm0.8\\
\mathrm{Praesepe} & 601 & 700 & 185.0 & 9.2\pm0.4 & 14.7\pm0.6 & 35.0\pm1.4 & 21.9\pm0.9 & 24.5\pm1.0\\
\mathrm{RSG\ 7} & 67 & 70 & 419.3 & 10.5\pm0.1 & 15.0\pm0.1 & 35.7\pm0.2 & 22.4\pm0.1 & 25.0\pm0.2\\
\mathrm{RSG\ 8} & 341 & 17 & 474.6 & 17.1\pm1.0 & 23.7\pm1.5 & 56.4\pm3.6 & 35.4\pm2.2 & 39.5\pm2.5\\
\mathrm{Roslund\ 5} & 191 & 97 & 540.5 & 12.5\pm1.5 & 17.3\pm2.4 & 41.2\pm5.7 & 25.8\pm3.6 & 28.8\pm4.0\\
\mathrm{Stephenson\ 1} & 262 & 46 & 358.6 & 10.3\pm0.9 & 14.7\pm1.2 & 35.0\pm2.9 & 21.9\pm1.8 & 24.5\pm2.0\\
\mathrm{Stock\ 1} & 137 & 470 & 405.6 & 10.4\pm0.6 & 14.9\pm0.7 & 35.5\pm1.7 & 22.2\pm1.0 & 24.8\pm1.2\\
\mathrm{Stock\ 12} & 121 & 112 & 437.8 & 9.7\pm0.5 & 13.9\pm0.8 & 33.1\pm1.9 & 20.7\pm1.2 & 23.2\pm1.3\\
\mathrm{Stock\ 23} & 105 & 94 & 606.9 & 20.8\pm4.3 & 28.2\pm5.5 & 67.1\pm13.1 & 42.1\pm8.2 & 47.0\pm9.2\\
\mathrm{UBC\ 19} & 42 & 7 & 399.7 & 26.3\pm4.4 & 37.7\pm6.8 & 89.8\pm16.2 & 56.3\pm10.1 & 62.8\pm11.3\\
\mathrm{UBC\ 31} & 260 & 11 & 365.0 & 14.0\pm1.7 & 20.0\pm2.7 & 47.6\pm6.4 & 29.9\pm4.0 & 33.3\pm4.5\\
\mathrm{UBC\ 31\ gp1} & 58 & 11 & 339.1 & 12.0\pm2.3 & 17.9\pm3.0 & 42.6\pm7.1 & 26.7\pm4.5 & 29.8\pm5.0\\
\mathrm{UBC\ 31\ gp2} & 185 & 10 & 381.4 & 17.1\pm2.4 & 24.5\pm3.6 & 58.3\pm8.6 & 36.6\pm5.4 & 40.8\pm6.0\\
\mathrm{UBC\ 7} & 191 & 40 & 278.1 & 10.3\pm1.3 & 15.4\pm2.0 & 36.7\pm4.8 & 23.0\pm3.0 & 25.7\pm3.3\\
\mathrm{UPK\ 82} & 57 & 81 & 542.2 & 6.8\pm2.2 & 9.4\pm3.0 & 22.4\pm7.1 & 14.0\pm4.5 & 15.7\pm5.0\\

	\enddata
	\begin{tablecomments}{
	    $M_{\rm cl}$ is the total mass of each star cluster.
	    The age of the cluster is obtained from \citet{pang2022a}, and was derived from PARSEC isochrone fitting. Dist. is the distance after distance correction from \citet{pang2022a} via Bayesian method.
	    Columns 5--6 present the binary fraction of each cluster.
	    $f^{q>0.4}_{bin}$ is the binary fraction computed using Equation~\ref{eq:binary_frac}. $f^{q>0.4}_{binc}$ is the binary fraction after completeness correction, as described in Sections~\ref{sec:complt}. $f^{tot}_{k07}$, $f^{tot}_{f05}$ and $f^{tot}_{uni}$ are the total binary fractions, assuming mass ratio distributions from \citet{kouwenhoven_primordial_2007,fisher_what_2005}, as well as a uniform mass ratio distribution (Section~\ref{sec:total_bf}).
	    }
    \end{tablecomments}
\end{deluxetable*}


\subsection{Completeness Correction}\label{sec:complt}

Our method for identifying unresolved binary candidates is subject to bias due to the limited angular resolution of Gaia in resolving binary systems. We apply a minimum angular separation for two-parameter solutions of 0.6\,arcsec \citep{lindegren_gaia_2021} to our sample of member stars. When the separation between the two binary components is below 0.6\,arcsec, the system is considered as an unresolved binary in Gaia data. 

The projected separation between the two components of a binary system at a given moment in time depends on the semi-major axis, the eccentricity, the spatial orientation of the orbit, and the orbital phase. For an ensemble of randomly-oriented binary systems with identical semi-major axis $a$ and eccentricity $e$, the average projected separation is
\begin{equation}
\overline{\rho} = \frac{\pi}{4} a \left( 1 + \frac{e^2}{2} \right) \quad ,
\end{equation}
which ranges from $\overline{\rho} \approx 0.79 a$ for circular orbits to $\overline{\rho} \approx 1.18 a$ for near-parabolic orbits \citep[see section 4.3.1 in][for detalis]{kouwenhoven2006}. In this study, we assume that the binary separation is approximately equal to the semi-major axis of the system. 

\begin{figure}[tb!]
\centering
\includegraphics[angle=0, width=0.35\textwidth]{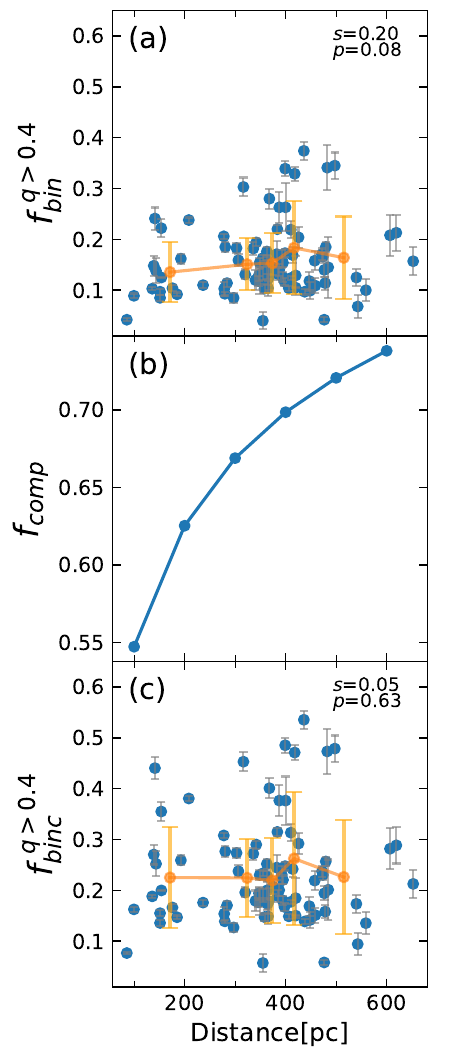}
    \caption{(a): The relation between binary fraction for $q>0.4$ identified in Section~\ref{sec:select} (blue dots) and the distance of the clusters. The error bar of $f^{q>0.4}_{bin}$ is estimated in Section~\ref{sec:err}. The orange dots are average values of cluster distance and $f^{q>0.4}_{bin}$ for all 17 clusters in each bin, with the standard deviation indicated with the error bar. (b): Dependence of the completeness fraction on distance. $f_{\rm comp}$ is estimated from Gaia's angular resolution limit at a given distance taken from Table~\ref{tab:bin_comp}. (c): Binary fraction after incompleteness correction $f^{q>0.4}_{binc}$ versus cluster distance. The error bar of $f^{q>0.4}_{\rm binc}$ is propagated from the error estimated in Section~\ref{sec:err}. The computation scheme of the orange dots is identical to that of (a).
    The quantity $s$ is the Spearman’s rank correlation coefficient, and $p$ is the probability of the null hypothesis (no correlation exist between two variables) of the correlation test. A $p$ value less than 0.1 means the null hypothesis is rejected.}
\label{fig:BF_comp}
\end{figure}

The identification of unresolved binaries through the CMD in Section~\ref{sec:select} is only able to recover those systems with semi-major axes smaller than the threshold value 0.6\,arcsec. However, binary systems with a separation larger than the minimum value are resolved as two single stars, which are not included in our estimation. This bias is particularly pronounced in nearby clusters, since a significant fraction of their binary systems is resolved. This effect is demonstrated in Figure~\ref{fig:BF_comp} (a), where the binary fraction $f^{q>0.4}_{bin}$ increases with cluster distance. Consequently, the binary fraction of nearby clusters is underestimated.

To estimate the completeness fraction of our identified binary candidates, we adopt a lognormal probability density function to describe the distribution of semi-major axis for FGK stars, taken from \citet{raghavan_survey_2010}, given that our target binary candidates are primarily FGK stars. The binary population in the Galactic field is primarily composed of binaries that originate from star clusters and stellar groups. While the filamentary young stellar groups are dispersing, their binary populations likely have a higher wide binary fraction than those of old star clusters, in which binary stars have been dynamically processed for a longer period. Therefore, the properties of the binary population in the field are the result of dynamical evolution in different environments: a combination of binaries originating from old disrupting clusters and young dispersing stellar groups. Both cluster types are well represented in our sample. We therefore adopt the distributions of \citet{raghavan_survey_2010} as an approximation for the binary stars in our sample. Note that using the distribution of \citet{raghavan_survey_2010} overestimates the wide binary population for dense clusters. Therefore, the corresponding completeness-corrected values will be treated as upper limits for the denser clusters in our sample. The lognormal distribution of semi-major axis $a$ has a mean value of 40 astronomical units (AU) and a dispersion of $\log (a/{\rm AU}) = 1.5$. The corresponding completeness fraction $f_{\rm comp}$ for binary systems of each respective cluster is obtained by integrating the distribution of $a$ up to the 0.6\,arcsec separation limit from Gaia, whose physical value (in AU) varies depending on the distance of clusters (as outlined in the column~3 of Table~\ref{tab:bin_comp}). As the cluster distance increases, the completeness fraction $f_{\rm comp}$ generally increases as well (as illustrated in Figure~\ref{fig:BF_comp} (b)), ranging from 0.55 to 0.75. In the nearest 100~pc neighboring region, the CMD identification method fails to detect approximately 45\% of the binary population in the star cluster. The binary fraction $f^{q>0.4}_{bin}$ for each cluster is adjusted by dividing the value $f^{q>0.4}_{bin}$ by the corresponding completeness fraction $f_{comp}$ at that distance (column~4 in Table~\ref{tab:bin_comp}). We group the target clusters into distance bins with a step size of 100~pc. Clusters in the same bin are assigned the same value of $f_{comp}$. The resulting corrected  binary fraction $f^{q>0.4}_{binc}$ is presented in the 6th column of Table~\ref{tab:bin_frac}. As shown in Figure~\ref{fig:BF_comp} (c), the dependence of binary fraction on distance is reduced after the correction, indicated by the Spearman's rank correlation coefficient, which changes from $s$ = 0.2 (a) to $s$ = 0.05 (c).

\begin{figure}[tb!]
\centering
\includegraphics[angle=0, width=0.45\textwidth]{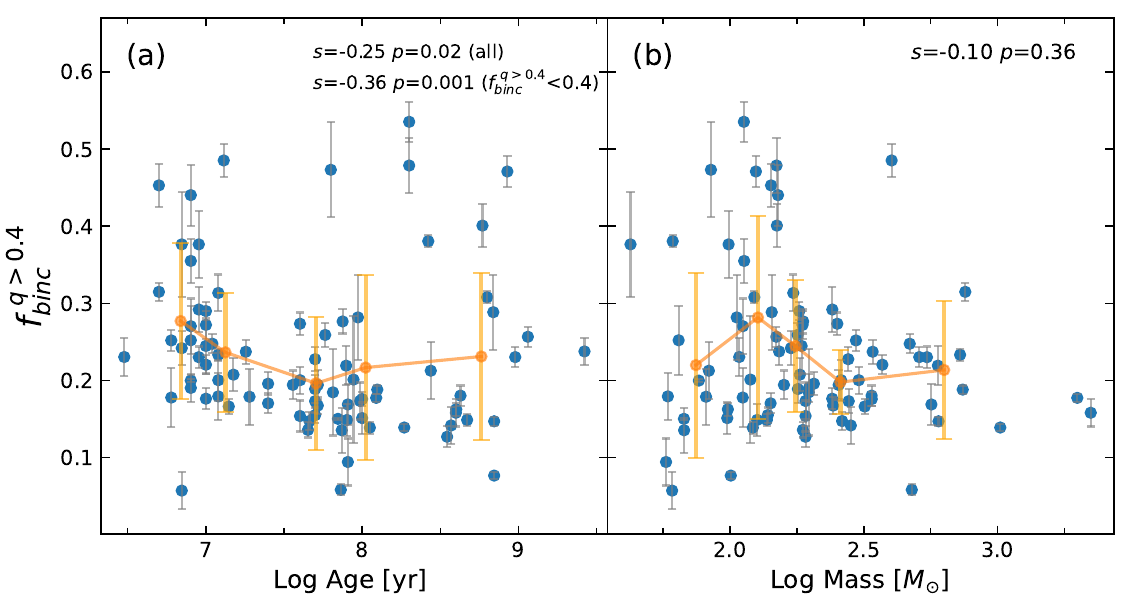}
    \caption{
   The relations between the completeness-corrected binary fraction for $q>0.4$ identified in Section~\ref{sec:complt} and the age of cluster (panel a) and the cluster total masses (panel b). The error bar of $f^{q>0.4}_{\rm binc}$ is propagated from the error estimated in Section~\ref{sec:err}. The orange dots represent the average values of the cluster distances and $f^{q>0.4}_{binc}$  for all 17 clusters in each bin, with the standard deviation indicated with error bars. The Spearman’s rank correlation coefficient $s$, and the probability of the null hypothesis $p$ are indicated in the upper right corner of each panel. The meanings of the quantities $s$ and $p$ are identical to those in Figure~\ref{fig:BF_comp}.
    }
\label{fig:BF_age_mass}
\end{figure}

\begin{deluxetable}{C C C C}
	\tablecaption{Binary separation limit for Gaia data at different distances.
	\label{tab:bin_comp}}
    \tabletypesize{\footnotesize}
	\tablehead{	    
	    \colhead{Distance (pc)}  & 
		\colhead{$\theta$ (arcsec)}  & 
        \colhead{$a$ (AU)}       & 
        \colhead{$f_{comp}$}    
		\\
	    \cline{1-4}
	    \colhead{(1)}   & 
	    \colhead{(2)}   & \colhead{(3)} & 
	    \colhead{(4)}   
		}
	\startdata
    100 & 0.6  & 60  & 0.55 \\
    200 & 0.6  & 120 & 0.63 \\
    300 & 0.6  & 180 & 0.67 \\
    400 & 0.6  & 240 & 0.70 \\
    500 & 0.6  & 300 & 0.72 \\
    600 & 0.6  & 360 & 0.74 \\
	\enddata
\tablecomments{$\theta$ is the minimum angular separation for two-parameter solutions in Gaia data. $a$ is the corresponding semi-major axis computed from $\theta$ for the binary system. $f_{\rm comp}$ is the completeness fraction for our identified binary candidates at different distances. }
\end{deluxetable}


\subsection{Total binary fraction}\label{sec:total_bf}

To estimate the total binary fraction, we apply an extrapolation method that involves using the binary fraction $f_{bin}^{q>0.4}$ for systems with mass ratios $q>0.4$ and applying it to binary systems with $q<0.4$. This is achieved through the use of a given mass-ratio distribution function $f(q)$. For extremely low-mass ratio systems, the companion may be a brown dwarf or planet, instead of a star; such systems are normally not considered as stellar binary systems. Therefore, the total binary fraction obtained in this manner is considered as an upper limit. \cite{kouwenhoven2009}, however, have shown that for the spectral types considered in our sample (FGK stars), the overestimation is small for the mass ratio distributions described below. Our approach is as follows.
(1) First, we make an assumption regarding the mass-ratio distribution function $f(q)$, based on the work by \citet{kouwenhoven_primordial_2007}, which takes the form of 
\begin{equation}
    f(q)\propto q^{-0.4\pm0.1} 
    \quad .
\end{equation}
This particular distribution is dominated by low-mass-ratio binary systems. The total binary fraction of the cluster is then estimated by multiplying the previously calculated completed binary fraction $f^{q>0.4}_{binc}$ (6th column in Table~\ref{tab:bin_frac}) with a factor $2.38\pm0.31$, which depends on this assumed $f(q)$ distribution. 
\begin{equation}
    f_{k07}^{tot}={(2.38\pm0.31)} f_{binc}^{q>0.4}
    \quad .
\end{equation}
 The error from the factor is propagated to the binary fraction, which is the highest among the three mass ratio profiles. Considering this mass ratio is obtained for young association, it may overestimate the general binary fraction for older clusters. Therefore, we will not use the total binary fraction for quantitative analysis. The resulting values are listed in column~7 of Table~\ref{tab:bin_frac}.

(2) In contrast to the mass-ratio distribution assumed in \citet{kouwenhoven_primordial_2007}, the distribution presented in \citet{fisher_what_2005} exhibits an increase in the frequency of high-mass-ratio binary systems, with a peak at around $q\sim0.9-1.0$. Over 70\% of binary stars have mass ratios greater than 0.4. Adopting the $f(q)$ distribution from \citet{fisher_what_2005}, we estimate the total binary fraction to be
\begin{equation}
    f_{f05}^{tot}={1.49}f_{binc}^{q>0.4} 
    \quad .
\end{equation}
 The values $f_{f05}^{tot}$ \citep{fisher_what_2005} are listed in column~8 of Table~\ref{tab:bin_frac}.   The coefficient is obtained from the number count of the histogram in \citet{fisher_what_2005}, for which no uncertainty is provided. We therefore do not compute the uncertainty for this mass ratio distribution. A similar procedure has been carried out by \citet{milone_acs_2012}.

(3) Assuming a uniform $f(q)$ for binary stars, the mass ratio remains constant across all values of $q$. Previous studies have shown that solar type stars in the field and star cluster follow roughly uniform distribution \citep{li_modeling_2020,torres_long-term_2021,offner_origin_2022}. In this scenario, the total binary fraction is given by
\begin{equation}
    f_{uni}^{tot}={1.67}f_{binc}^{q>0.4}
    \quad ,
\end{equation}
which value is listed in the last column of Table~\ref{tab:bin_frac}. The coefficient for the uniform distribution is exact, and no uncertainty is computed. Considering the under-estimated uncertainty of the total binary fraction, we do not use it for quantitative analysis in the remainder of this study. Elaborating upon the distribution of mass ratio in a more comprehensive manner goes beyond the scope of the current study. 


\section{Global binary properties}\label{sec:dependence}

According to Gaia angular resolution limit (Section~\ref{sec:complt}), the binary systems we have identified in this study are classified as intermediate to close binaries, with separations of $\leq$360~AU \citep{offner_origin_2022}. The survival of binary systems in clustered environments depends not only on their own energy but also on the energy and frequency of encounters \citep{parker_binaries_2014}. Therefore, the hardness of the binary system, which is defined by the ratio between its binding energy and the kinetic energy of the system's center of mass \citep{parker2012,reipurth_multiplicity_2014}, is of great importance. The kinetic energy of the binary system and encounter rates, and thus the hardness, are significantly affected by the stellar density. Using the current number of members and cluster size \citep[half-mass radius obtained from][]{pang2022a}, we estimate the hard-soft boundary $a_{hs}$ for our cluster sample using the equation (1) provided by \citet{reipurth_multiplicity_2014}. The value of $a_{hs}$ ranges from a few hundred AU up to a few thousand AU. However, due to mass loss of the cluster and observational incompleteness of members, this value of $a_{hs}$ is an upper limit, and it was likely much smaller at the time of birth of the cluster. Therefore, the binary sample we have studied in the 85 target clusters certainly includes both soft and hard binaries.


\subsection{Binary fraction vs. Age}\label{sec:bf_age}

The evolution of a cluster tends to lead to changes in the hardness of its constituent binary systems, with hard binaries becoming harder and soft binaries becoming softer \citep{heggie2003}. Soft binaries are likely to eventually disrupt as the cluster undergoes secular evolution, leading to a decline in the overall binary fraction as the cluster ages. Figure~\ref{fig:BF_age_mass} (a) depicts the relationship between the completeness corrected binary fraction $f_{binc}^{q>0.4}$ and cluster age, with a clear scatter observed across all ages. Despite the scatter among some clusters with $f_{binc}^{q>0.4}$ values above 40\%, the majority follows the trend of decreasing binary fraction with increasing age ($s$=-0.36). The small value of $p$ indicates that a correlation exist between cluster age and $f_{binc}^{q>0.4}$, although the Spearman's rank correlation coefficient $s$ indicates that the degree of correlation is modest. The scatter in the binary fraction is equally large at young ages ($\leq$10\,Myr) as it is at old ages ($\sim$1\,Gyr). Young stellar groups and clusters are ideal laboratories for studying primordial binary properties. The dispersion in the binary fraction may be caused by different binary initial conditions in different clustered environments during the star formation process. As wide and soft binaries are disrupted as clusters evolve, the total number of binaries declines. At the same time, single stars also continue to escape from the cluster and consequently the total member number decreases, thus artificially pushing up the binary fraction. Therefore, the scatter of binary fraction in old clusters provides insights into  the degree of dynamical disruption of clusters. 


\subsection{Binary fraction vs. Cluster mass}\label{sec:bf_mass}

In the Figure~\ref{fig:BF_age_mass} (b), we show the dependence of binary fraction on cluster mass. There exists an increasing trend in clusters with masses less than 200\,$\rm M_\odot$, where the binary fraction rises from 20\% to 30\% on average. In clusters with higher masses, the total binary fraction appears to be unaffected by the cluster mass, and the scatter is also reduced. There is statistical indication that a higher binary fraction in lower-mass clusters than higher-mass counterparts in our sample, 
which may be attributed to the lower stellar density in low-mass clusters. The collision rates and mean kinetic energy of colliding stars are both lower in such environments \citep{sollima_fraction_2010}. However, the Spearman's rank correlation coefficient, $s=-0.10$, suggests a weak correlation between the binary fraction and the cluster mass, similar to the results of globular clusters \citep{milone_acs_2012}.


\subsection{Binary fraction vs. Stellar density}\label{sec:bf_density}

\begin{figure*}[tb!]
\centering
\includegraphics[angle=0, width=1.0\textwidth]{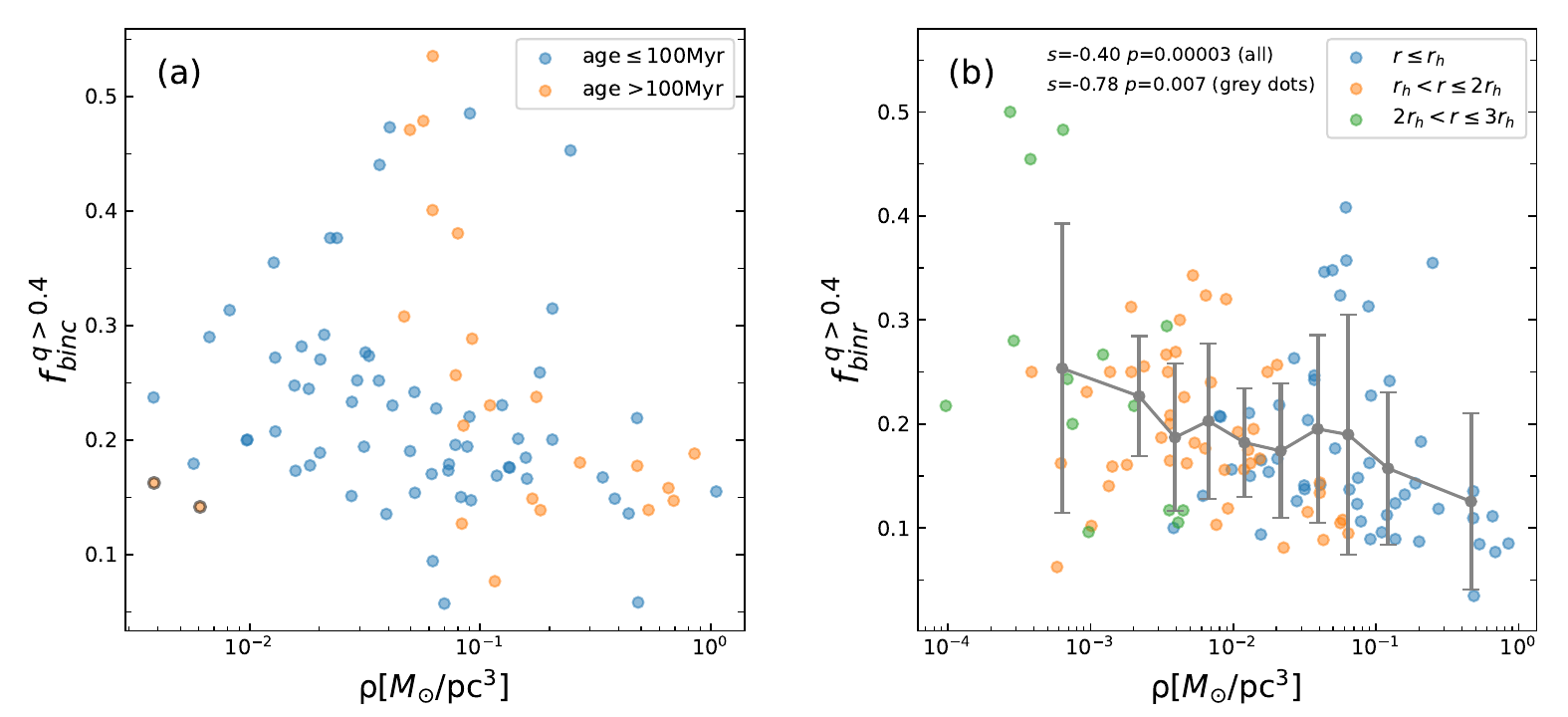}
    \caption{
     The correlation between the completeness corrected binary fraction for $q>0.4$ computed in Section~\ref{sec:complt} and the stellar density within the half-mass radius (panel a). The blue dots are clusters with ages less than 100\,Myr, while the orange dots are older than 100\,Myr. Two orange dots with black circles are outliers: the disrupted cluster Group\,X which fragments into two over-densities, and the tidal-tail cluster Mamajek\,4 with very low central density.  In panel (b), we display the radial binary fraction versus the radial stellar density in each cluster, for regions with binary number more than 5. The blue dots refer to binary fraction and stellar density for stars within half-mass radius \rh; orange dots for stars between \rh{} and $2r_{\rm h}$; green dots for stars between $2r_{\rm h}$ and $3r_{\rm h}$. The grey dots are mean value of each bin with the same number of clusters. The vertical error bar is the standard deviation of the binary fraction in each bin. 
      The meanings of the quantities $s$ and $p$ are identical to those in Figure~\ref{fig:BF_comp}.
     }
\label{fig:BF_density}
\end{figure*}

To explore the link between binary fraction and stellar density in clusters, we investigate the correlation between the corrected binary fraction $f^{q>0.4}_{binc}$ and the central stellar density ($r\leq r_{\rm h}$) in Figure~\ref{fig:BF_density} (a). However, we do not observe an obvious dependence of binary fraction on central stellar density for any of the clusters. 
On the contrary, \citet{niu_binary_2020} has found that binary fraction increases with higher stellar density. Their 12 open clusters are all older than 100\,Myr with high stellar density within \rh{} ($\rho$ is in the range of 1--7\,$M_\odot\,pc^{-3}$). 
We highlight clusters older than 100\,Myr in Figure~\ref{fig:BF_density} (a) (orange dots), and conversely observe an opposite trend that binary fraction decreases with higher stellar density. No evident trend is observed in the clusters younger than 100\,Myr (blue dots in Figure~\ref{fig:BF_density} (a)).

In order to study the binary fraction dependence on stellar density in the same cluster, we conduct an analysis of the radial distribution of the binary fraction in three annular regions within our sample of clusters, defined by radial distances of $r \leq r_{\rm h}$, $r_{\rm h} < r \leq 2r_{\rm h}$, and $2r_{\rm h} < r\leq 3r_{\rm h}$. To ensure statistically reliable outcomes, we restrict our examination to annuli comprising at least five binary candidates. We plot the mean value of binary fraction in each density bin (with an identical number of clusters) in dark grey dots in Figure~\ref{fig:BF_density} (b). The vertical error bar is the standard deviation of the binary fraction in each density bin. From the outskirt to the central core region, the mean binary fraction of each bin (dark grey dots) decreases from 25\% to 15\%. This reveals a noticeable decline in the binary fraction with increasing density in the same clustered environment, considering the relatively strong correlation indicated by the Spearman’s rank correlation coefficient with a value of $s=-0.78$. Among the clusters and groups in our sample, 70\% is younger than 100\,Myr. The decline of the binary fraction toward high density regions is evidence of early binary disruption in young stellar systems. 
This is the most robust relation compared to binary fraction with age or cluster mass, indicating that stellar density probably is a fundamental parameter determining binary evolution.

\section{Binary fraction in different environments}\label{sec:environ}

\subsection{Mean radial profile of binaries}\label{sec:morph} 

The 85 open clusters examined in this study have been classified by \citet{pang2022a} into four distinct types based on their morphology: filamentary, fractal, halo, and tidal-tail. These four types correspond to different clustered environments. The filamentary and fractal types are both less than 100~Myr and exhibit elongated filament shapes or fractal structures, respectively. They are formed along the filaments of molecular clouds with relatively low star formation efficiency \citep{kruijssen2012}. The density of these clusters or groups, as seen in Figures~7 and~8 of \citet{pang2022a}, has the lowest density among the four types, making them more favorable to binary survival \citep{portegies_zwart_star_2004,parker_evolution_2011}. Due to their youth and low-density environments, the binary properties of the young filamentary and fractal groups are most similar to those of primordial binaries.

On the contrary, halo-type clusters are characterized by a high-density core and are the most massive and dense clusters in our sample. The high interaction rate within this dense environment disrupts binary evolution. At the same time, violent stellar encounters in the dense environment can foster tidal capture of companions and the formation of close binaries.
This mechanism is more likely to occur for massive stars \citep{torniamenti2021}.

Finally, the tidal-tail clusters represent dynamically evolved clustered environments, where the binary stars that have survived the cluster's evolution have already undergone a few or dozen relaxation times. The current stellar density within tidal-tail clusters ranges from intermediate to low density. The binary members in these clusters are likely on their way to leaving the cluster, and we expect that the observed binary properties of this type of cluster should be somewhat similar to those of field binaries.

The results of our analysis reveal that filamentary and fractal types have the highest global binary fraction, with an average of 23.6\%$\pm$9.2\% and 23.2\%$\pm$7.2\%, respectively. Tidal-tail clusters or groups have a slightly lower average binary fraction of 20.8\%$\pm$9.5\%, ranking second in terms of completeness corrected binary fraction $f^{q>0.4}_{binc}$. On the other hand, the halo-type clusters in our study exhibit the lowest binary fraction among the four types, with a mean of 14.8\%$\pm$9.5\%. These findings are consistent with theoretical predictions that low-density clustered environments are more favorable for binary survival and therefore have a higher binary fraction \citep{portegies_zwart_star_2001,parker_binaries_2014}.

\begin{figure}[tb!]
\centering
\includegraphics[angle=0, width=0.45\textwidth]{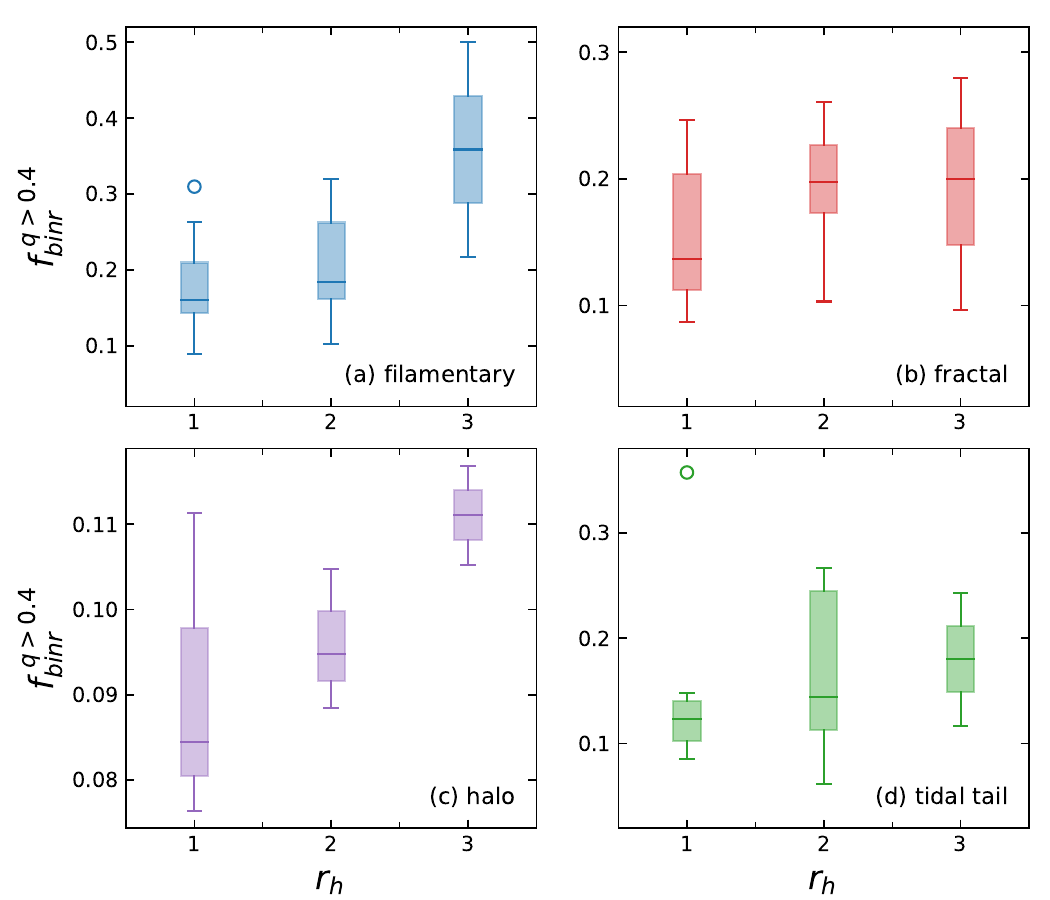}
    \caption{
    The dependence of radial binary fraction on cluster-centric distance in the unit of half-mass radius \rh{} in four types of clusters: filamentary (panel a), fractal (panel b), halo (panel c), and tidal-tail (panel d).  The three colored rectangles for each type indicate the value for stars within \rh{}, between \rh{} and $2r_{\rm h}$, and between $2r_{\rm h}$ and $3r_{\rm h}$. The colored rectangles indicate the inner quartile range (IQR, 75 percentile minus 25 percentile). The median value is indicated with a horizontal line. The upper and lower whiskers of each colored rectangle mark the maximum and minimum values within 1.5IQR. Clusters outside the 1.5IQR (whiskers) are outliers indicated as open circles: Collinder\,69 (blue circle) and Collinder\,350 (green cicle). 
    }
\label{fig:BF_type}
\end{figure}

In order to investigate the influence of the local stellar density on binary fraction within each cluster or group, we conduct an analysis of the radial distribution of binary stars in different clustered environments, as shown in Figure~\ref{fig:BF_type}. We define regions in each cluster or group based on the dynamical scale, specifically the half-mass radius \rh{}, with three annuli considered: $r \leqslant r_{\rm h}$, $r_{\rm h} < r \leqslant 2r_{\rm h}$, and $2r_{\rm h} < r \leqslant 3r_{\rm h}$. Our results indicate that the radial binary fraction for each of the four morphological types agrees with the global mean value, and that all types exhibit a decline in binary fraction towards the center of the cluster or group, where the density is highest. These findings are consistent with those presented in  Figure~\ref{fig:BF_density}, and suggest that local stellar density plays a critical role in shaping binary evolution in a different clustered environment. 

 As pointed out by \citet{parker_evolution_2011}, 
binary systems in subvirial and substructured star clusters will experience fast rate of dynamical processing in local dense regions, which are formed during the phase of cool collapse. The highly-energetic dynamical interactions that occur in dense regions can efficiently disrupt soft binaries. Moreover, even in the expanding stellar groups,  i.e., the filamentary and fractal type clusters, local over-density regions might dynamically process primordial binaries.
Consequently, the binary populations in these clusters may have processed, and their properties may thus be different from those at birth. As our current data do not allow us to quantify the dynamical history of the targeted stellar groups/clusters, we cannot draw conclusive remarks on this issue.

\subsection{Binary Radial Profile for Individual Clusters}\label{sec:morph_indi} 

\begin{figure*}[tb!]
\centering
\includegraphics[angle=0, width=1.\textwidth]{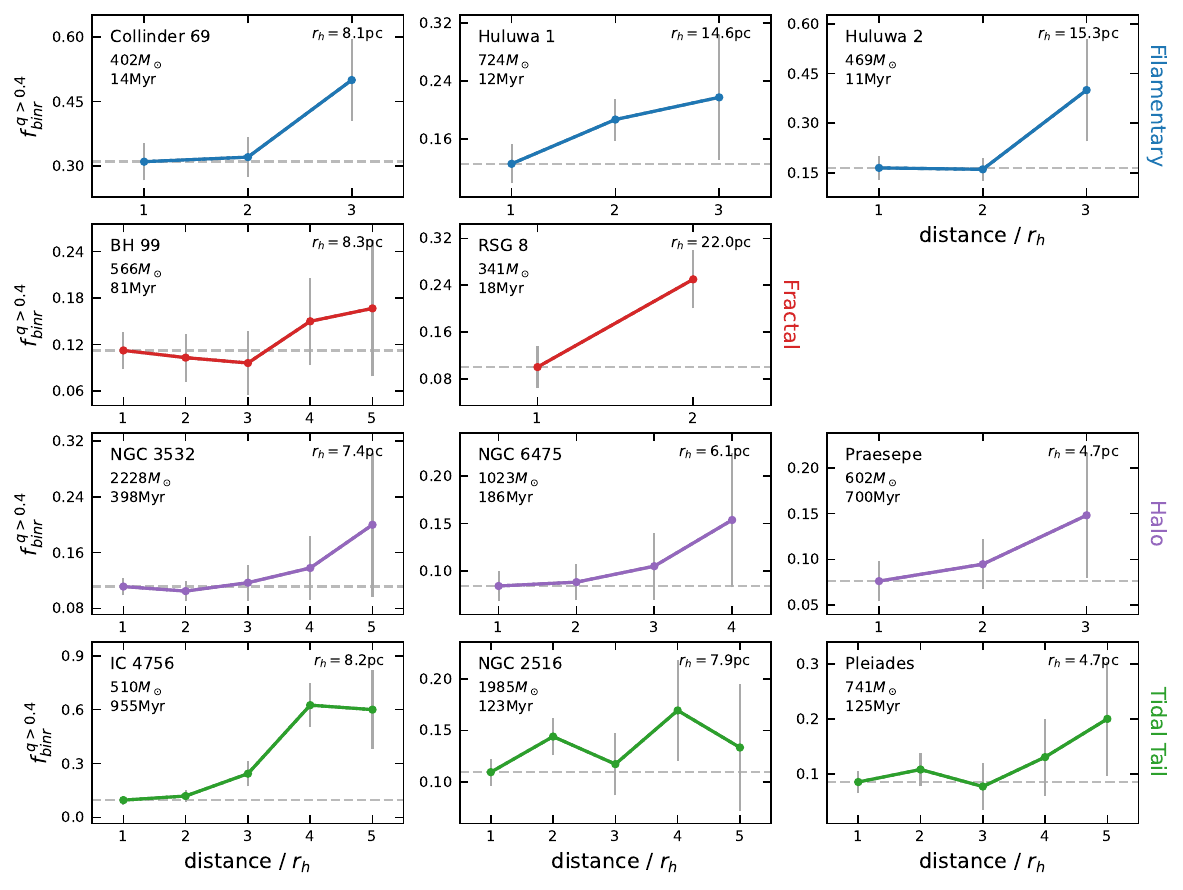}
    \caption{The radial dependence of the binary fraction on the cluster-centric distance for the representative clusters of each of the four morphological types. These clusters are selected based on the numbers of binary and single star are both greater than 28. The cluster-centric distance is in the unit of half-mass radius \rh{}. The types of cluster are marked with different colours. The horizontal dashed line indicates the value of $f^{q>0.4}_{binr}$ for the first bin (within \rh{}). The grey bars indicate the errors of binary fraction of each annulus, which is the standard deviation of the sampling distribution of binary fraction. The color coding is identical to that of Figure~\ref{fig:BF_type}. 
    }
\label{fig:bf_radial}
\end{figure*}

For demonstration purposes, we select clusters from each morphological type that contain more than 28 binary candidates (total members $>56$) and plot their radial binary profile in Figure~\ref{fig:bf_radial}. The width of the annulus is half-mass radius \rh{}. As expected from the global relation shown in Figure~\ref{fig:BF_type} for all morphological types, the radial binary fraction is lowest inside the half-mass radius and gradually increases towards the outer part of the cluster. This is a result of the disruptive effects of dynamical evolution, which causes a significant fraction of soft binaries to be destroyed in the cluster center due to encounters.  We witness the early binary disruption in open clusters. 

A similar trend is observed in the star cluster NGC\,1818, in which the binary fraction declines toward the core
\citep{li_binary_2013,grijs2013, geller_different_2015}. For a cluster born with soft or wide binaries, the early evolution of the binary population is dominated by disruption, which decreases the overall binary frequency and establishes a decreasing trend in binary fraction toward the cluster core within approximately a crossing time. 
The simulations of \citet{geller_consequences_2013} have shown that a binary fraction evolves over time from one that decreases toward the core to a bimodal distribution, and eventually to a distribution that increases only toward the core, which is observed in dynamically old star clusters \citep{milone_acs_2012,geller_different_2015}.
This evolution process occurs at the two-body relaxation timescale. Two of the clusters in our sample (LP\,2429 and NGC\,6774) are older than 1\,Gyr in our sample are already unbound, disrupted, and form tidal tails. Both singles and binaries are leaving the system. Their radial profile of binaries follows the general trend we find in Figure~\ref{fig:bf_radial}. We do not find evidence of a peak of binary fraction in the cores of these two old clusters. Due to the persistent tidal field on the Galaxy, open clusters typically become unbound and dissolve within 100-200\,Myr \citep{dinnbier2020a}. Binaries often  escape the cluster before they reach the center. Therefore, a bound system is the pre-condition to maintain binary stars and nurture central binary evolution. 

\subsection{Mass segregation in different environments}\label{sec:mass-seg}

As clusters evolve, they experience two-body relaxation, leading to mass segregation, which is more significant in dynamically-evolved systems. Since binaries are typically more massive than single stars, they tend to sink to the cluster center faster due to the inverse proportionality between the  segregation timescale and stellar mass \citep{pang2013}. The higher-mass binaries undergo dynamical friction and mass segregation processes at a faster rate than low-mass binaries, and therefore, although all binaries are subject to dynamical disruption early on, the higher-mass binaries begin increasing their core binary frequency more quickly \citep{geller_different_2015}. 

\begin{figure*}[tb!]
\centering
\includegraphics[angle=0, width=1.\textwidth]{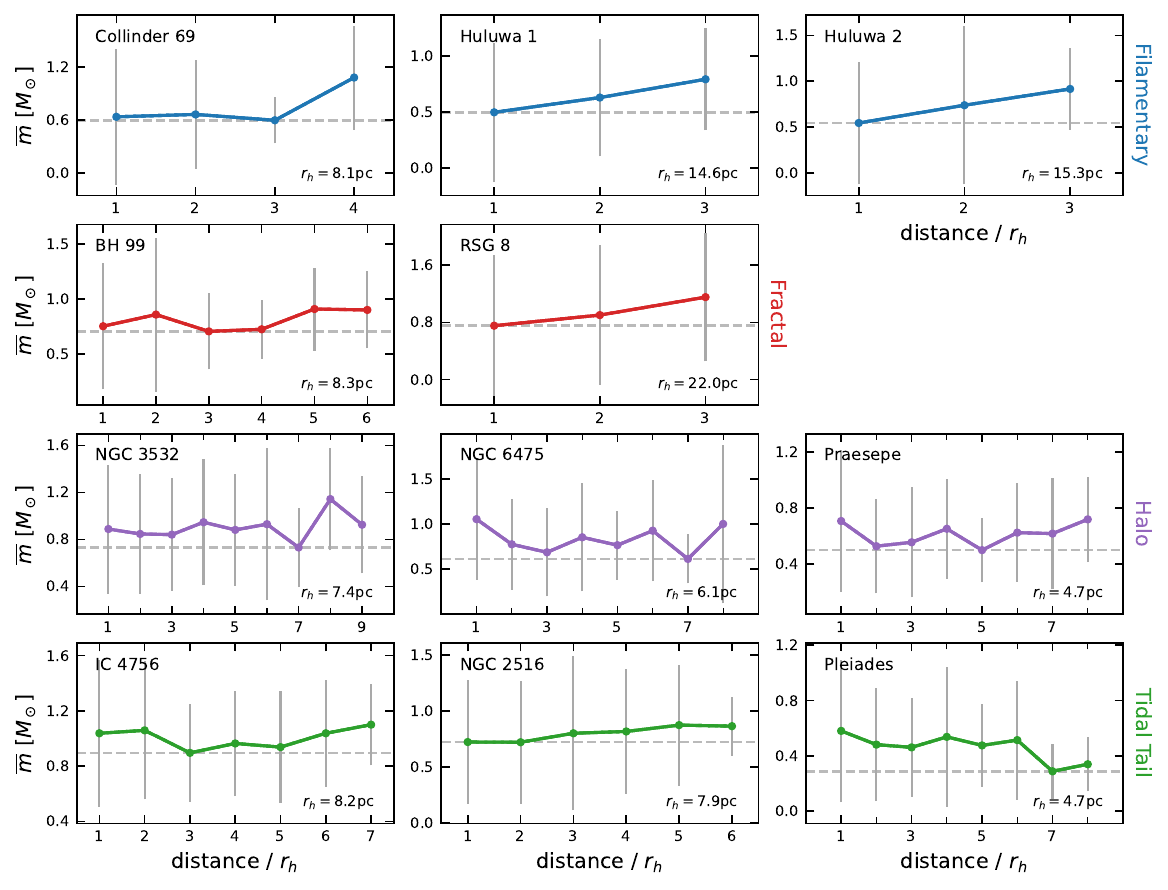}
    \caption{
    The dependence of mean mass on cluster-centric distance for the representative cluster of each of the four types shown in Figure~\ref{fig:bf_radial}. The size of each annulus is the half-mass radius \rh{} of each cluster. Each dot represents the mean mass in the annulus, and the error bar indicates the standard deviation in each bin. The horizontal dashed line indicates the smallest mean mass value among all bins. The color coding is identical to that of Figure~\ref{fig:bf_radial}. 
    }
\label{fig:mass_seg_2rh}
\end{figure*}

We study the mass segregation of four types of clusters or groups by calculating the average stellar mass in different annuli with a size of half-mass radius \rh{}, as shown in Figure~\ref{fig:mass_seg_2rh}. We detect no mass segregation in filamentary and fractal types. Interestingly, this types of young stellar groups seem to exhibit reverse segregation, where higher-mass stars are located further away from the center. However, due to their irregular and extended morphology, the median position of members might not accurately represent the group center. On the other hand, the most significant mass segregation is observed in halo cluster  NGC\,6475, while a less pronounced but still observable amount of mass segregation is found in halo cluster Praesepe, and tidal-tail clusters IC\,4756 and Pleiades, which is consistent with findings from previous studies of the same clusters \citep{roser2019,Lodieu19,ye2021}. 

Since the filamentary, fractal stellar groups and the tidal-tail clusters are extended in space, the approach of using spherical annuli may not be suitable to quantify the spatial mass distribution. 
In \citet{pang2021a,pang2022a}, the 3D morphology of open clusters is quantified by a fitted ellipsoid. The direction of the semi-major axis $d_a$ is considered as the elongated direction of star cluster. 
Because stellar distances from inverting the parallax generates an apparent stretching along the line of sight,  
the 3D $X$, $Y$, $Z$ coordinates of each star from \citet{pang2022a} were corrected. \citet{pang2020,pang2021a,pang2022a}
mitigated this problem via a Bayesian approach developed by \citet{carrera2019}, which adopted two priors: a normal distribution for the individual distances to the cluster stars and an exponentially decreasing profile for the distances to field stars \citep{bailer2015}. The error for the corrected distance ranges from 0.3--6\,pc, depending on the orientation of the primordial shape of the cluster relative to the line-of-sight \citep[the primodial shape of the cluster may be sperical or elongated along the line-of-sight;][]{pang2021a}. 

\begin{figure*}[tb!]
\centering
\includegraphics[angle=0, width=1.\textwidth]{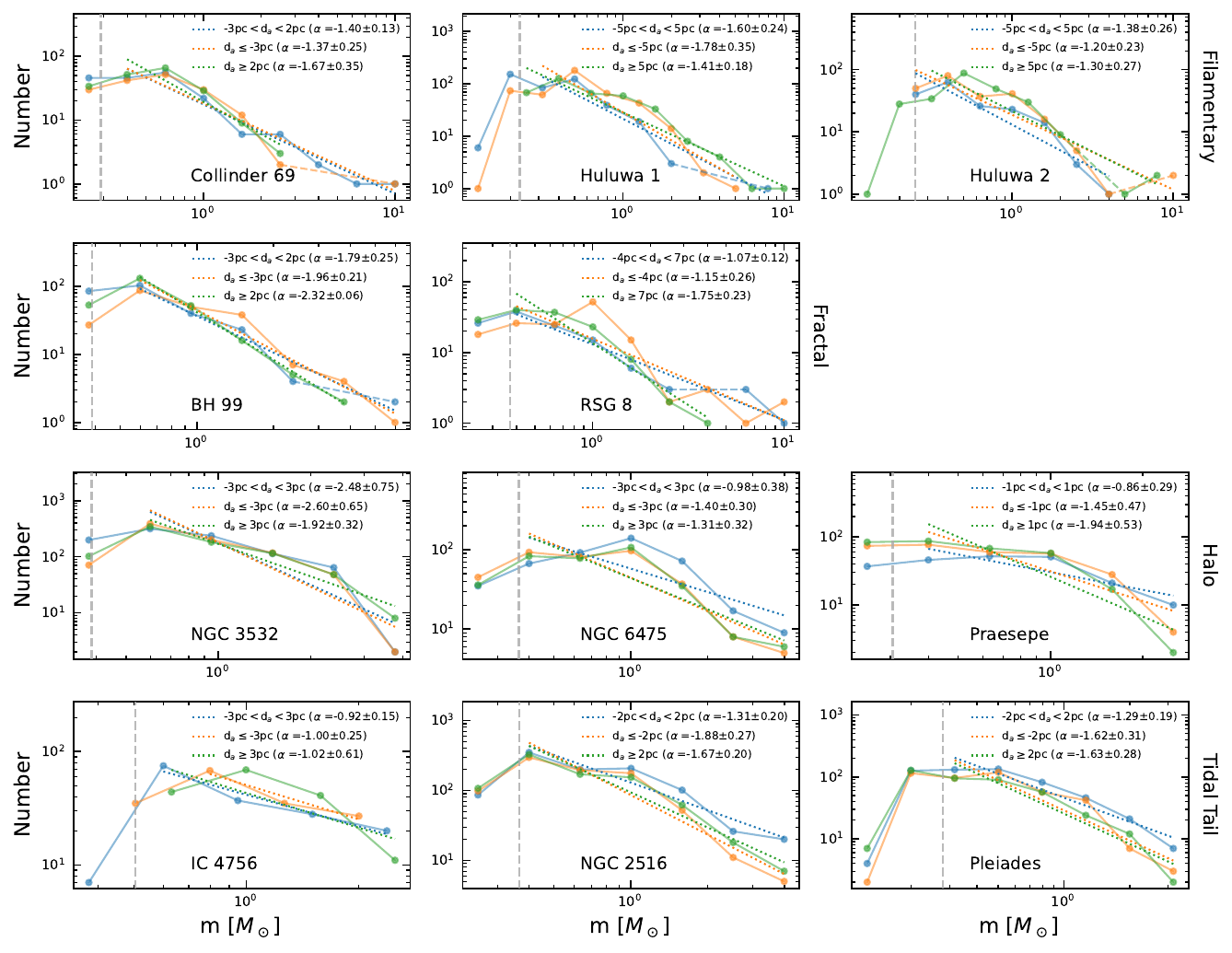}
    \caption{
    The stellar mass function for three regions along the semi-major axis $d_a$ of the best-fitting ellipsoid of each representative cluster shown in Figure~\ref{fig:bf_radial}. The 3D position of each star is transformed into the coordinate system spanned by $d_a$, $d_b$, and $d_c$, which are the the semi-major, semi-mediate, semi-minor axes of the bested fitted ellipsoid from \citet{pang2022a}. The blue curve represents the mass function in the inner region, a few pc away from the center in the  $d_a$ direction. the orange and green curves are mass functions in the region on the left (negative $d_a$ values) and right side (positive $d_a$ values) of the inner region along $d_a$ direction. The best fit to the mass function of each region is indicated with the dotted colored lines, whose slopes, $\alpha$ and corresponding uncertainties are indicated in the upper-right corner of each panel. The vertical grey dashed line is the Gaia completeness limit at the distance of the cluster.  
    }
\label{fig:mass_seg_da}
\end{figure*}

We convert the 3D position of each star from Galactic coordinates into the Cartesian coordinates of $d_a$ (semi-major axis of the best fitted ellipsoid), $d_b$ (semi-mediate axis of the best fitted ellipsoid), and $d_c$ (semi-minor axis of the best fitted ellipsoid), with the origin of the coordinate located at the cluster center (median position of all members). In Figure~\ref{fig:mass_seg_da}, we show the stellar mass function of three regions along the direction of semi-major axis ($d_a$): inner region with $|d_a|$ less than a few pc; two regions on the left (negative $d_a$ values) and right side (positive $d_a$ values)  of the inner region.  The size of each region is selected in order to assure that the number of stars in each region is similar, so that they are equally statistically significant. We find hints for flatter slopes of the mass function, $\alpha$ (indicated in the upper-right corner of each panel in Figure~\ref{fig:mass_seg_da}), in the inner region for the clusters BH\,99, RSG\,8 (fractal-type), Praesepe (halo-type), and  NGC\,2516 (tidal-tail type), than in the outer regions, providing evidence for mass segregation. Considering the uncertainty of $\alpha$, the observed mass segregation is not significant. However, in the filamentary clusters, no mass segregation is observed. Instead, in this clusters, some high-mass stars are located beyond the central region of the cluster, which could be considered as inverse mass segregation.  

\section{Velocity dispersions}\label{sec:vel-disp}

To quantitatively analyze the internal dynamical properties of our target clusters, we compute the dispersion of the proper motion (PM) of cluster members within and outside \rh{}, with the goal of identifying differences in the kinematic features of stars inside and outside \rh{}. We divide the cluster members into two groups based on their location: within or beyond \rh{}.

We utilize the Markov Chain Monte Carlo (MCMC) method to obtain the optimal values and associated uncertainties for the PM dispersions. We model the likelihood function of the proper motion distribution ($\mu_\alpha \cos\delta$ and $\mu_\delta$) by two Gaussian functions \citep[Equations 1--3 in][]{pang2018}: one for the cluster members and the other for the field component (which contributes only 5\% to the overall distribution). 

\begin{figure}[tb!]
\centering
\includegraphics[angle=0, width=0.5\textwidth]{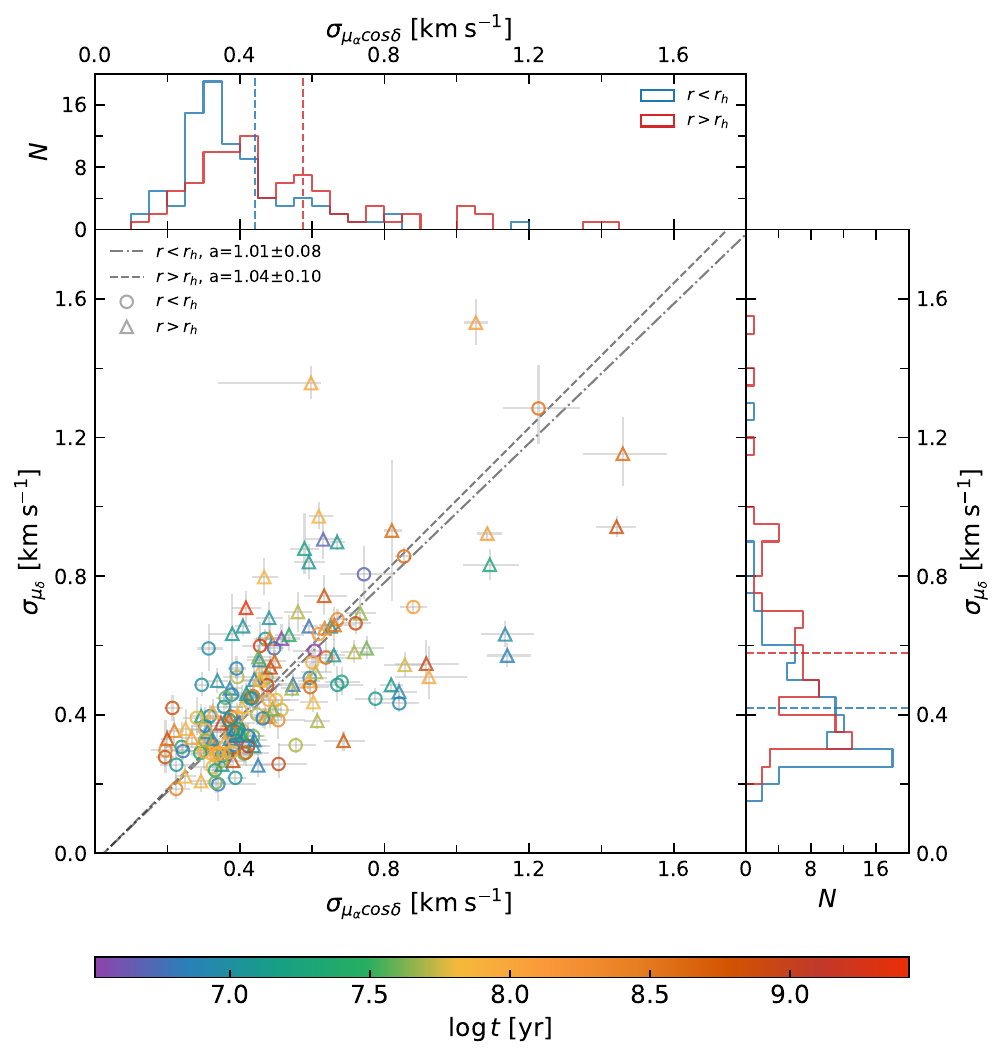}
    \caption{
    The distribution of the dispersion of the 1D tangential velocity, along right ascension and declination. The velocity dispersion value is computed using the MCMC method (see Section~\ref{sec:vel-disp}). Circles indicate dispersion values obtained from stars within the half-mass radius \rh{} and the triangles are for stars outside \rh{}. The color of symbols is scaled with the cluster age (color bar). The dashed and dashed-dotted lines are the linear fits to the correlation of $\sigma_{\mu_{\alpha}}$ and $\sigma_{\mu_{\alpha}\cos\delta}$ for region outside and inside \rh{} respectively, with errors taken into consideration. The histograms in the upper and right panels represent the value of $\sigma_{\mu_{\alpha}\cos\delta}$ and $\sigma_{\mu_{\alpha}}$ for stars inside \rh{} (blue) and outside \rh{} (red). The blue and red dashed line in the upper and right panels denote the mean value of $\sigma_{\mu_{\alpha}\cos\delta}$ and $\sigma_{\mu_{\alpha}}$ for stars inside and outside \rh{}, respectively.
    }
\label{fig:pm_disp}
\end{figure}

Figure~\ref{fig:pm_disp} displays the one-dimensional velocity dispersion for tangential velocity along right ascension and declination, which is converted from PM velocity dispersions based on the distance of each cluster. The open circles correspond to values obtained from stars located inside \rh{}, while triangles represent those outside \rh{}. Taking into account the uncertainty, the velocity dispersions generally exhibit isotropy along the two tangential velocity components. The mean one-dimensional dispersion for members outside \rh{} is approximately 0.14--0.16~${\rm km\,s^{-1}}$ larger than that inside \rh{}.
As demonstrated in Figures~\ref{fig:BF_density}, \ref{fig:BF_type}, and~\ref{fig:bf_radial}, the binary fraction is higher outside \rh{}, indicating that the effects of binaries on proper motion measurements should be considered.

\subsection{Binary's influence on velocity dispersion}\label{sec:infl}

Within the typical timespan of an observational campaign, a single star follows a linear and uniform trajectory in the sky. The point at which its light is concentrated, known as the photocenter, corresponds to the location of the star. In the case of an unresolved binary system, both stars orbit around their common barycenter, which does not necessarily coincide with the location of the photocenter. When the mass ratio is low, the location of the photocenter of the system is near the position of the primary star and therefore significantly deviates from the location of the barycenter. When Gaia observes the displacement of the photocenter of an unresolved binary system, it exhibits an astrometric wobble or anomaly \citep{wang_astrometric_2022}, which affects the accuracy of proper motion measurements and increases their uncertainty. The primary astrometric data of Gaia's current releases are obtained by treating all objects as single stars \citep{lindegren_gaia_2021}. Binary candidacy is indicated by a high renormalized unit weight error (RUWE). 
We compute the mean value of RUWE for binary candidates and single members in our 85 target clusters. The mean RUWE for binaries is 1.6, compared to 1.1 of singles, which tends to confirm the candidacy of the identified binaries. 

We carry out a simulation to estimate the bias in the 1D tangential velocity $\Delta V_t$ along the declination due to the deviation of photocenter from barycenter induced by the unresolved binary systems. The main motivation for the binary simulation is to estimate the influence of unresolved binary systems on the measurements of the proper motion, and to quantify the uncertainty in the tangential velocity dispersion measurements due to unresolved binary systems. In the case of an unresolved binary, Gaia measures the astrometry of the photocenter instead of that of the barycenter. The cluster velocity dispersion should be determined using the barycenter of the cluster. The photocentric motion will contribute a small fraction to the total velocity dispersion, and the measured velocity dispersion should thus be corrected for this effect.

In this simulation, we adopt a mass of 1\,$M_\odot$ for the primary star. Given the mass ratio $q$, the mass of the secondary is then $q M_\odot$. Light emitted by the secondary is considered in this simulation. The position of the binary system is randomly assigned within a sphere of unit radius. We do not make any assumption about the distance to the cluster. For a binary system with certain mass ratio and separation, we calculate the 2D projection of the nonlinear motion of the photocenter on the plane of the sky. The velocity of the photocenter of this  unresolved binary only depends on the mass of the companion $q M_\odot$, the semi-major axis $a$ and the total $G$-band luminosity of the binary. Hence no assumption of distance is required.

For simplicity, the eccentricity of the binary system is assumed to be zero, and the barycenter of the system is assumed to have zero proper motion. We consider the astrometric observational baseline as three years \citep{gaia_collaboration_gaia_2022}. A total of 36 random observing times are generated within this period of three years. We compute the semi-major axis of the orbit of the photocenter using the mass of the companion. The mass-luminosity relation \citep{pecaut_intrinsic_2013} of MS stars is used in this step. The motion of the target system's photocenter in the orbital plane of target system is fit with a linear model. The velocity of binary stars is converted into the Cartesian sky coordinate system using a rotational matrix which is produced by the Gram–Schmidt process \citep{Schmidt1907}. The matrix determines the orientation of the orbital plane and the initial phase of velocity components. The velocity is projected onto the direction of declination. The velocity difference of the photocenter and the barycenter is the bias triggered by binary systems. Since barycenter is set to have zero PM, the magnitude of the projected value is the 1D tangential velocity bias from binary systems. This procedure is repeated 400 times to obtain the average bias for a given semi-major axis $a$ distribution \citep{raghavan_survey_2010} and mass ratio $q$ distribution \citep{fisher_what_2005}. 

\begin{figure}[tb!]
\centering
\includegraphics[angle=0, width=0.5\textwidth]{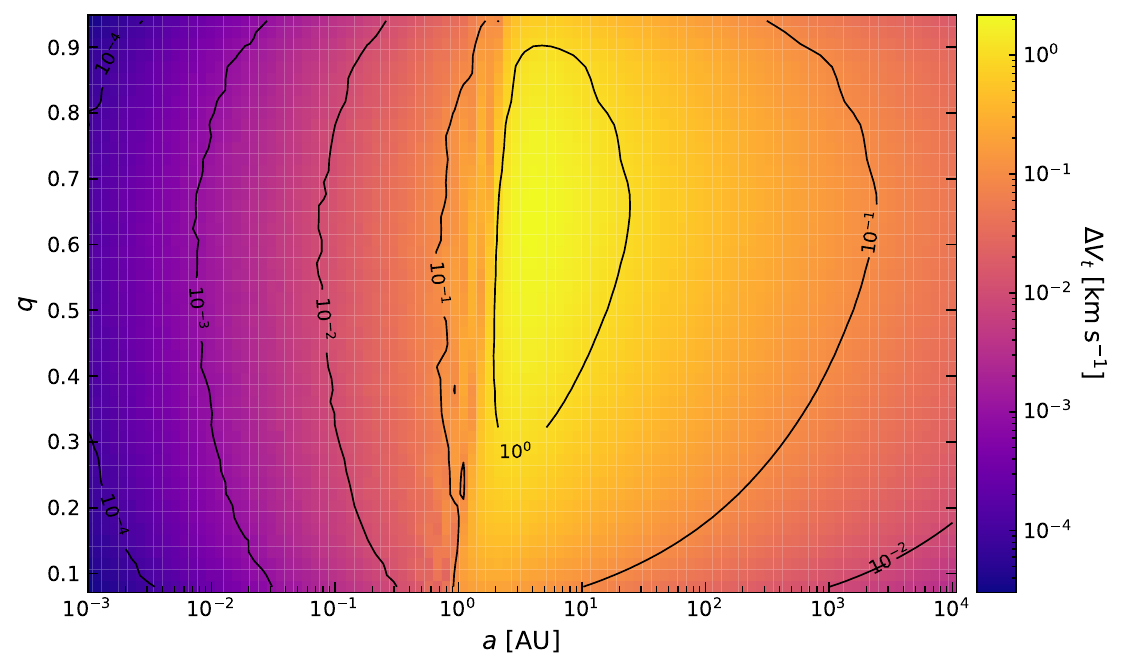}
    \caption{
    The dependence of the bias in the 1D tangential velocity along declination direction on the binary semi-major axis $a$ and the mass ratio $q$. When calculating the bias, we assume a primary mass of 1\,$M_\odot$. The color bar indicates the average 1D tangential velocity bias along the declination direction. 
    }
\label{fig:vt_er}
\end{figure}

The results of the simulation are shown in Figure~\ref{fig:vt_er}. For binary systems with longer period (larger $a$), i.e., the right quadrant of Figure~\ref{fig:vt_er}, the bias $\Delta V_t$ increases for smaller semi-major axes ($a<100$\,AU) and larger companion masses ($q$=0.3--0.9). 
However, short-period binary systems (left quadrant in Figure~\ref{fig:vt_er} for $a<1$\,AU) generally generate a smaller bias in the tangential velocity when the period becomes smaller. 
When $q=1$, the barycenter of the binary system coicides with the photocenter, thus leading to zero bias for $\Delta V_t$.

 Our identification approach recover binary with semi-major axis $a$ less than 360~AU (Section~\ref{sec:complt}), accordingly these kind of unresolved binary systems induce an additional error 0.1--1\,$\rm km\,s^{-1}$ (Figure~\ref{fig:vt_er}) in the measurement of the tangential velocity (obtained from the proper motion) larger than that of single stars.
We obtain the 1D tangential velocity dispersion of binary stars and single members respectively via the same MCMC method. The dispersion of unresolved binary stars is 0.1--0.5\,$\rm km\,s^{-1}$ larger than that of the single stars, shown in Figure~\ref{fig:pm_disp}. However, this larger value might not be intrinsic. It can be caused by the bias induced by the unresolved binary systems.
As shown in Figures~\ref{fig:BF_density}, \ref{fig:BF_type} and~\ref{fig:bf_radial}, the unresolved binary fraction is higher at larger distances from the cluster center. Therefore, the unresolved binary fraction outside half-mass radius is generally higher than that inside the half-mass radius. A higher velocity dispersion outside half-mass radius is likely a consequence of the higher fraction of unresolved binary stars, which induce the larger uncertainty in the proper motion measurement. We propose this as one possibility to explain the larger velocity dispersion outside the half-mass radius.
 
Therefore, based on the current accuracy of Gaia PM measurements, the higher dispersion of stars outside \rh{} (relative to those inside), might not be a pure dynamical signature. We are unable to draw a reliable conclusion about the internal kinematics of star clusters if the effects induced by the unresolved binaries are not properly considered and subtracted. 

In the simulation described above, the binary population is not placed in a clustered environment, since our goal is not to investigate the dynamical evolution of binaries in star clusters. The latter would require $N$-body simulations, which is beyond the scope of our current study.


\section{Cross-Match with previous works}\label{sec:cross}

Numerous investigations have been conducted to determine the binary fraction of star clusters using various methods. In this section we compare the total binary fraction obtained in this work to that of  previous studies, and investigate the reasons for any discrepancies that may arise.

In the study by \citet{jadhav_high_2021}, the binary fraction for $q>0.6$ was estimated for 23 open clusters using a similar method via CMD. However, unlike the current study, they used a larger magnitude range of  $M_G$=1--10 mag in Gaia absolute $G$ magnitude to include a wider range of stars. We find that their binary fractions for all of the seven overlapped open clusters (IC\,4756, NGC\,2422, NGC\,2516, NGC\,2547, NGC\,3532, NGC\,6774, Pleiades) are consistent with the total binary fractions estimated in this study after completeness correction.
Using infrared photometry,  \citet{malofeeva_unresolved_2022-1} identified binary candidates for the Pleiades cluster and estimated a binary fraction of $0.54\pm0.11$. In this study, a total binary fraction of 0.28--0.45 is derived for the Pleiades cluster, assuming different mass ratio distributions, which agrees with the value obtained in \citet{malofeeva_unresolved_2022-1}. 
The recent study of \citet{malofeeva2023} updated the unresolved binary fraction for Pleiades to $0.55\pm0.02$, and obtained a binary fraction $0.45\pm0.03$ for Praesepe. Both values are higher than the binary fraction derived in this work. 

Thirty clusters in our sample are also described in the study of  \citet{donada_multiplicity_2023}. The latter study also carried out distance-dependent completeness corrections for the binary fraction. In general, our results for the binary fraction (after completeness correction) are in agreement with their results, except for three clusters: Alessi\,9, NGC\,1901, and NGC\,3228. For these clusters we find higher binary fractions than \citet{donada_multiplicity_2023}. This difference might be attributed to the different lower limit for the mass ratio ($q>0.6$) adopted when calculating the binary fraction. 

Six clusters in our sample appear in the study of \citet{cordoni_photometric_2023}, which obtained binary fractions for $q>0.6$ using a similar method in the CMD as this work. All binary fraction of the matched clusters (Pleiades, NGC\,2442, NGC\,2516, NGC\,2547, NGC\,3532 and Stock\,12) are consistent with our results within the uncertainties. 

In the case of NGC\,3532, \citet{li_modeling_2020} derived a binary fraction of 0.267, which agrees well with the total binary fraction estimated in this study assuming a uniform mass ratio distribution. 
Finally, \citet{sollima_fraction_2010} estimated binary fractions for five open clusters ranging from 36\% to 70\%, with NGC\,2516 overlapping with this study. In our study we derive a total binary fraction of 27\%--42\% for NGC\,2516, which is consistent with the binary fraction of $65.5\pm24.3\%$ from \citet{sollima_fraction_2007}.

The utilization of different methodologies to determine binary fractions yields both advantages and disadvantages. In compiling binary fractions for research purposes, it is imperative to consider the types of binaries that are being identified and the feasibility of the identification method, such as the range of semi-major axis and mass ratio, to ensure homogeneity in the resulting binary data.


\section{Summary}\label{sec:summary}

Using the member catalogs from \citet{pang2021a,pang2021b,pang2022a} and \citet{li2021}, we identify binary systems within 85 open clusters based on their locations in the color magnitude diagram. Specifically, we select systems with mass ratios greater than $q_0=0.4$ as candidates. We define the binary region in the color-magnitude diagram as the region with an upper boundary at an absolute $G$ magnitude ($M_G$) equal to 4\,mag and a lower boundary at stellar mass equal to 0.5 $M_\odot$ (see Section~\ref{sec:identify}). The binary fraction, $f^{q>0.4}_{bin}$, is computed as the number of binary systems relative to the total number of binary systems and single stars within each cluster.

The uncertainty of the binary fraction is estimated using Monte Carlo simulations. In this process, we generate artificial stars based on the photometric uncertainties and recompute the binary fraction using these simulated stars. We then compare this value to the observed binary fraction, $f^{q>0.4}_{bin}$, and quantify the difference as the associated uncertainty.

To ensure an accurate measurement of the binary fraction, $f^{q>0.4}_{bin}$, for each cluster, we adopt completeness correction to account for the limitation of Gaia's angular resolution of 0.6~arcsec at different distances. To achieve this, we estimate the completeness fraction $f_{comp}$ using the log-normal semi-major axis distribution from \citet{raghavan_survey_2010}. The completeness corrected binary fraction $f^{q>0.4}_{binc}$ is then computed as the ratio of $f^{q>0.4}_{bin}$ to $f_{comp}$ within a given distance bin. 

We consider three mass ratio distributions to extrapolate the $f^{q>0.4}_{binc}$ to obtain the total binary fraction:  \citet{kouwenhoven_primordial_2007} ($f^{tot}_{k07}$), a uniform distribution ($f^{tot}_{uni}$) and \citet{fisher_what_2005} ($f^{tot}_{f05}$). 
Below, we summarise our main findings regarding the binary candidates in the targeted clusters.

\begin{itemize}
    \item The relationship between the completeness-corrected binary fraction $f_{binc}^{q>0.4}$ and cluster age exhibits a considerable dispersion across all age ranges. By excluding clusters or groups with $f_{binc}^{q>0.4}>0.4$, we observe a decreasing trend in the overall binary fraction with increasing cluster age.
    
    \item The completeness-corrected binary fraction $f_{binc}^{q>0.4}$ exhibits a positive correlation with cluster total mass for clusters with $M_{cl}<200\,M_\odot$. However, for clusters with $M_{cl}>200\,M_\odot$, $f_{binc}^{q>0.4}$ appears to be independent of the cluster mass.
    
    \item The local stellar density plays a crucial role in the dynamical binary evolution of binary systems. Our investigation has revealed a robust correlation between increasing stellar density within a cluster and a decrease in the binary fraction. We find this radial trend of binary fraction in various morphological types of clusters, which have different density environments. In the filamentary and fractal stellar groups or clusters with the lowest density, we observe the highest binary fraction $f_{binc}^{q>0.4}$ with mean values of 23.6\%$\pm$9.2\% and 23.2\%$\pm$7.2\%, respectively. The tidal-tail clusters exhibit a mean $f_{binc}^{q>0.4}$ of approximately 20.8\%$\pm$9.5\%, while the densest halo-type clusters have the smallest binary fraction of 14.8\%$\pm$9.5\%.
    
    \item We investigate the radial distribution of binary fraction for representative clusters belonging to four distinct types. We find that clusters of all four types showed a consistent trend in which the binary fraction is observed to be lowest within the half-mass radius (\rh{}), and increases toward the larger cluster-centric radius. This observation provides compelling evidence that the binary population experiences early disruption in the dense regions of clusters.
    
    \item The analysis of a representative set of clusters belonging to four types of environments reveals a distinct pattern in the mean mass distribution. In particular, the filamentary groups exhibit inverse mass segregation, while the fractal-type clusters BH\,99 and RSG\,8, the halo-type cluster Praesepe, and the tidal-tail clusters NGC\,2516 display hints of mass segregation. However, considering the uncertainty, the observed mass segregation is not significant. Therefore, the observed mass segregation does not yet generate a global effect inside the target clusters. Furthermore, the tidal-tail clusters and some expanding filamentary stellar groups were found to be in a disrupted state \citep{pang2022a}, which was not conducive to the evolution of the binary population. Binary systems in such clusters are instead likely to follow the expanding flow and escape, and are thus spared from internal dynamical processing. 
    
    \item By assuming a primary mass of 1\,$M_\odot$, we estimate that the bias induced by unresolved binary systems in 1D tangential velocity along declination direction ranges from 0.1 to 1~$\rm km\,s^{-1}$. Consequently, we cannot draw definitive conclusions regarding the internal kinematics of the clusters based on the present Gaia proper motion data.
    
\end{itemize}

Our study is limited by the angular resolution of the Gaia data. Moreover, the assumption of a primary mass of 1\,$M_\odot$ in our simulation (Section~\ref{sec:vel-disp}) is a simplification made to estimate the bias in the 1D velocity dispersion induced by unresolved binary systems. Further studies using a wider range of primary masses are needed to address these limitations. Our findings demonstrate that it is worthwhile to continue investigations of the relationship between the binary fraction and environmental properties such as cluster age, mass, and stellar density, with higher-resolution data. Future spectroscopic and astrometric data from space-based and ground-based telescopes, will provide a better understanding of the formation and evolution of binary systems in open clusters.

\acknowledgments
We thank the anonymous referee for providing helpful comments and suggestions that helped to improve this paper.
We thank Yuqian Li for help with the ellipsoid coordinate transformation, Danchen Wang for help with the velocity dispersion computation, and Teng Zhang for constructive discussions. 
Xiaoying Pang acknowledges the financial support of the
 National Natural Science Foundation of China through grants 12173029 and 12233013, Natural Science Foundation of Jiangsu Province (No. BK20200252), and the research development fund of Xi'an Jiaotong-Liverpool University (RDF-18--02--32). 
M.B.N.K. acknowledges support from the National Natural Science Foundation of China (grant 11573004) and Xi'an Jiaotong-Liverpool University (grant RDF-SP-93). 

This work made use of data from the European Space Agency (ESA) mission {\it Gaia} 
(\url{https://www.cosmos.esa.int/gaia}), processed by the {\it Gaia} Data Processing 
and Analysis Consortium (DPAC, \url{https://www.cosmos.esa.int/web/gaia/dpac/consortium}). This study also made use of 
the SIMBAD database and the VizieR catalogue access tool, both operated at CDS, Strasbourg, France.


\software{  \texttt{Astropy} \citep{astropy2013,astropy2018,astropy2022}, 
            \texttt{SciPy} \citep{millman2011},
            \texttt{TOPCAT} \citep{taylor2005}, and 
            \textsc{StarGO} \citep{yuan2018}.
}
\clearpage
\bibliography{main}
\bibliographystyle{aasjournal}

\end{document}